\documentclass[11pt]{article}

\usepackage{amsmath,mathtools,amssymb,color,epsfig,cite}
\usepackage{amsfonts}
\usepackage{graphicx}
\usepackage{set	space}
\usepackage{mathrsfs}
\usepackage{tikz}
\usetikzlibrary{calc,decorations.markings}
\usetikzlibrary{arrows,decorations.pathmorphing}
\usepackage{makecell}
\usepackage{multirow}

\makeatletter
\newsavebox{\@brx}
\newcommand{\llangle}[1][]{\savebox{\@brx}{\(\m@th{#1\langle}\)}
  \mathopen{\copy\@brx\kern-0.5\wd\@brx\usebox{\@brx}}}
\newcommand{\rrangle}[1][]{\savebox{\@brx}{\(\m@th{#1\rangle}\)}
  \mathclose{\copy\@brx\kern-0.5\wd\@brx\usebox{\@brx}}}
\makeatother

\usepackage{amsfonts}

\newcommand{\hoch}[1]{$\, ^{#1}$}

\makeatletter
\@addtoreset{equation}{section}
\makeatother

\newcommand{\be}{\begin{equation}}
\newcommand{\ee}{\end{equation}}
\newcommand{\bea}{\setlength\arraycolsep{2pt} \begin{eqnarray}}
\newcommand{\eea}{\end{eqnarray}}

\def\biblio{\bibliographystyle{utphys}\bibliography{ANEC}}

\def\0{{\sst{(0)}}}
\def\1{{\sst{(1)}}}
\def\2{{\sst{(2)}}}
\def\3{{\sst{(3)}}}
\def\4{{\sst{(4)}}}
\def\5{{\sst{(5)}}}
\def\6{{\sst{(6)}}}
\def\7{{\sst{(7)}}}
\def\8{{\sst{(8)}}}
\def\sst#1{{\scriptscriptstyle #1}}

\def\ep{{\epsilon}}
\def\del{{\partial}}

\def\C{\mathcal{C}}

\def\D{\Delta}

\begin{document}
\def\biblio{}

\vspace*{15pt}
\begin{center}
{\Large {\bf Lorentzian CFT 3-point functions in momentum space}}

\vspace{25pt}
{\bf Teresa Bautista\hoch{1} and Hadi Godazgar\hoch{2}}

\vspace{20pt}

{\it Max-Planck-Institut f\"ur Gravitationsphysik (Albert-Einstein-Institut), \\
M\"ühlenberg 1, D-14476 Potsdam, Germany.}

 \vspace{25pt}
January 8, 2020

\vspace{30pt}

\underline{ABSTRACT}
\end{center}

\noindent
In a conformal field theory, two and three-point functions of scalar operators and conserved currents are completely determined, up to constants, by conformal invariance. 
The expressions for these correlators in Euclidean signature are long known in position space, and were fully worked out in recent years in momentum space.
In Lorentzian signature, the position-space correlators 
 simply follow from the Euclidean ones by means of the $i\epsilon$ prescription. In this paper, we compute the Lorentzian correlators in momentum space and in arbitrary dimensions for three scalar operators by means of a formal Wick rotation. We explain how tensorial three-point correlators can be obtained and, in particular, compute the correlator with  two identical scalars and one energy-momentum tensor. As an application, we  
show that expectation values of the ANEC operator simplify in this approach.
 \noindent

\thispagestyle{empty}

\vfill
E-mails: \hoch{1}teresa.bautista@aei.mpg.de, \hoch{2}hadi.godazgar@aei.mpg.de

\pagebreak
\tableofcontents
\vspace{10mm}
\rule{400pt}{0.03pt}
\section{Introduction}
\label{sec:int}

Much work on conformal field theories, and quantum field theories in general, has focused on the Euclidean theory rather than the Lorentzian theory. In the former setting, correlation functions are simpler and obey satisfying properties, for example they are symmetric under permutations of operators or they are analytic when points are non-coincident. Furthermore, one can always `return' to the Lorentzian theory provided that certain requirements are satisfied via the Osterwalder-Schrader reconstruction theorem \cite{Haag:1992hx}.   However, it has recently been shown that a lot of mileage can be gained by instead thinking of the theory in the Lorentzian setting. While Wightman functions do not share the simplicity of their Euclidean cousins, their richer structure encodes important information about the theory. One such powerful property is causality, which has  been recently used to, for example, prove the average null energy condition (ANEC) \cite{Hartman:2016lgu} or in the analytic bootstrap program, see for example Refs.~\cite{Komargodski:2012ek, Fitzpatrick:2012yx,Hartman:2015lfa,Li:2017lmh,Costa:2017twz}, which among other things has lead to the Lorentzian inversion formula for CFTs \cite{Caron-Huot:2017vep}. 

Despite the recent interest in Lorentzian CFTs (see for example Refs.~\cite{Belin:2019mnx,Kologlu:2019bco} among many others), an expression of the correlation functions in momentum space beyond 2 points has hitherto been lacking. Even the Euclidean correlators in momentum space have only been fully  worked out relatively recently \cite{Coriano:2013jba, BMS:imp, BMS:renoms, Coriano:2017mux, BMS:renomtj, BMS:renomsjt, Coriano:2018bsy}, despite the need for them mainly from applications to cosmology in the context of which they were partially studied \cite{Maldacena:2011nz, Creminelli:2012ed} (see \cite{Sleight:2019hfp} for more recent cosmological applications).   

It is well known that conformal symmetry places restrictions on the form of correlators in a conformal field theory, fixing the 2-point functions, up to a normalisation, and 3-point functions, up to constants that are part of the CFT data. Solving the constraints on correlators from conformal symmetry in Euclidean signature has been done in position \cite{Polyakov:1970xd, Schreier:1971um, OP} and momentum space \cite{BMS:imp, BMS:renomsjt}. The complication in momentum space is that the special conformal Ward identity is a second order differential equation of generalised hypergeometric type~\cite{BMS:imp}, whereas in position space the solutions can be shown to be derivatives of powers of separation distances between the points at which operators are inserted. Directly Fourier transforming the position space correlators, while manageable for the scalar 3-point function \cite{BMS:imp}, is unwieldy for tensorial correlators. 
 
Given the recent attention to Lorentzian CFTs and the fact that it is often convenient to work in momentum space, in this paper we study 3-point functions of Lorentzian CFTs in momentum space. While these correlators can in principle be found by Fourier transforming the Lorentzian correlators in position space, which is already more difficult than the analogous Fourier transform in Euclidean space, we instead show that these correlators can be derived from the Euclidean expressions by a careful Wick rotation. The advantage of working in Lorentzian space is that the $i \epsilon$ prescription precludes coincident singularities, which seems to imply that renormalisation is not required. While our results for the scalar 3-point function are consistent with this claim (see section \ref{sec:div}), we leave a thorough analysis of this issue for future work. 

As an application of our results, we revisit the expectation values of the ANEC operator on the Hofman-Maldacena states \cite{HofMal} produced by scalar operators, which were calculated using correlators in position space. While this result is not new, it is illustrative of the fact that such calculations are  much more natural in momentum space where interesting features are not obscured. For example, it becomes clearer that the Hofman-Maldacena quantities are an expectation value. We believe that this perspective will also be indispensable in attempts to understand the implications of ANEC away from criticality. 

We begin, in section \ref{sec:pre}, by outlining the general procedure for Wick rotating from Euclidean to Lorentzian correlators. We use the scalar 2-point function as an example to elucidate the approach, which has applicability beyond the subject of interest in this paper. In section \ref{sec:scam}, it is shown that the scalar 3-point function can be bootstrapped from the 2-point function results and given in terms of an integral over an auxiliary momentum; as a check, in appendix \ref{sec:scalar-FT} we arrive at the same result in four dimensions by Fourier transforming the 3-point function in position space.  We also present the scalar 3-point function as an integral over three Bessel functions, section \ref{sec:tbe}, by Wick rotating the analogous Euclidean 3-point function given in terms of an integral of three $K$-Bessel functions. In the Lorentzian case, the integral includes both Bessel and modified Bessel functions. We discuss the finiteness of the scalar 3-point function in section \ref{sec:div} and show that renormalisation is not required. In section \ref{sec:check}, we check the equivalence of the two obtained expressions for the 3-point function for a particular example. In section \ref{sec:ten}, we give a general prescription for determining tensorial correlators from scalar correlators and apply this to find the correlation function of the energy-momentum tensor and two identical scalar operators. In particular, this example is considered not only because it is simple enough that the general prescription can be illustrated clearly, but also on account of its application in section \ref{sec:Hofmal}. As a demonstration of the utility of Lorentzian correlators in momentum space, we calculate the expectation value of the ANEC operator in a state produced by a scalar operator. In appendix \ref{sec:not}, we set out our notations and conventions. 

\section{2-point function}
\label{sec:pre}

We consider the 2-point function of scalar operators. This provides a useful playground to explore the relation between Euclidean and Lorentzian correlators in position and momentum space. For time-ordered 2-point function or 2-point functions of higher-spin operators see Ref.~\cite{Gillioz:2018mto}.

The Euclidean 2-point function of scalar operators $\mathcal{O}_1$ and $\mathcal{O}_2$ with dimensions $\Delta_1$, $\Delta_2$ is given by 
\begin{equation}
 \label{2ptE}
G_{\scriptscriptstyle{E}}(x) \equiv \langle \mathcal{O}_1(x) \mathcal{O}_2(0)  \rangle_{\scriptscriptstyle{E}} = \frac{ \delta_{\Delta_1,\Delta_2}}{x^{2 \Delta}}\,,
\end{equation}
where $\Delta\equiv\Delta_1=\Delta_2$; we have set the normalisation constant to be one and we have used translation invariance to set one of the positions to the origin. The 2-point function is non-zero only if the two operators have the same conformal dimension-- we assume this in the expressions that follow and henceforth drop the Kronecker $\delta$.\footnote{If the operators are characterised by other quantum numbers, they must also both have the same number in order for the 2-point function to be non-trivial. We also assume this henceforth.}

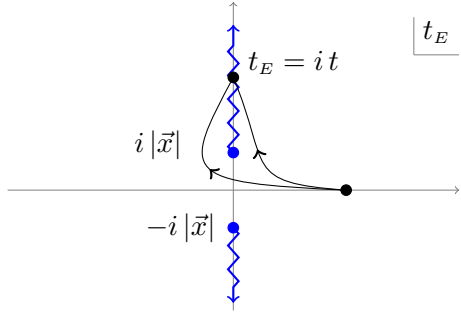
\begin{figure}
\centering
\begin{tikzpicture}[decoration={markings,
mark=at position 0.58 with {\arrow[line width=1pt]{>}},
}
]
\draw[help lines,->] (-3,0) -- (3,0) coordinate (xaxis);
\draw[help lines,->] (0,-1.6) -- (0,2.5) coordinate (yaxis);
\draw[color=blue,fill=blue] (0,0.5) circle (2pt);
\draw[color=blue,fill=blue] (0,-0.5) circle (2pt);
\draw[color=blue,line width=0.8pt,->,decorate,decoration={zigzag,amplitude=2,segment length=9,post length=4}] (0,0.5) -- (0,2.2);
\draw[color=blue,line width=0.8pt,->,decorate,decoration={zigzag,amplitude=2,segment length=9,post length=4}] (0,-0.5) -- (0,-1.5);
\draw[fill] (1.5,0) circle (2pt);
\draw[fill] (0,1.5) circle (2pt);
\draw[postaction=decorate] (1.5,0) .. controls (0.1,.1) and (0.5,0.3) .. (0,1.5);
\draw[postaction=decorate] (1.5,0) .. controls (-0.7,.1) and (-0.7,0.2) .. (0,1.5);
\draw[help lines] (2.4,1.8) -- (3,1.8);
\draw[help lines] (2.4,1.8) -- (2.4,2.3);
\node at (-1,0.6) {$i\, |\vec{x}|$};
\node at (-0.7,-0.5) {$-i\,|\vec{x}|$};
\node at (2.7,2.1) {$t_{\scriptscriptstyle{E}}$};
\node at (0.8,1.7) {$t_{\scriptscriptstyle{E}}=i\,t$};
\end{tikzpicture}
\caption{The two different Wick rotations from Euclidean to Lorentzian time, corresponding to passing either to the right or the left of the branch cut, define the two possible Wightman 2-point functions. The contour on the right corresponds to $\langle \mathcal{O}(x)\mathcal{O}(0)\rangle$ and the contour on the left to $\langle \mathcal{O}(0)\mathcal{O}(x)\rangle$.} \label{fig:branch-cut-2point}
\end{figure}

The Lorentzian  2-point function follows from performing a Wick rotation to Lorentzian time $t_{\scriptscriptstyle{E}} = i t$ in the Euclidean correlator \eqref{2ptE}. Since the correlator has a branch cut in the $t_{\scriptscriptstyle{E}}$ plane starting at $i|\vec{x}|$ and running upwards and from $-i\infty$ until $-i|\vec{x}|$, the Wick rotation is ambiguous for $|t|\geq|\vec{x}|$: $t$ can be defined to be either on the right or on the left of the cut (see figure \ref{fig:branch-cut-2point}). This ambiguity can be resolved by adding either a positive or a negative infinitesimal real part to $t_{\scriptscriptstyle{E}}$,  $t_{\scriptscriptstyle{E}} = i (t \pm i \epsilon)$.  Each one of these two options defines then a different Wightman function, and the so-called $i\epsilon$ prescription dictates that the imaginary part of the Lorentzian time of the operator to the left is more negative than that of the operator to the right, \textit{viz.} 
\begin{equation} \label{2ptL}
G(x) \equiv \langle \mathcal{O}(x) \mathcal{O}(0) \rangle = \frac{1}{\big({- (t - i \, \epsilon)^2 + | \vec{x}|^2}\big)^{\Delta}}\,,
\end{equation}
where for the ordering given above $\epsilon >0.$ For the Wightman function with the operator $\mathcal{O}(0)$ to the left, we then require $\epsilon<0$.

An intuitive way to understand the $i\epsilon$ prescription is that given a particular ordering of (any number of) operators, the $i\epsilon$'s are chosen in such a way that
\begin{equation}
 \dots e^{i H (t_i - i \epsilon_i)} \,\mathcal{O}_i(\vec{x}_i) \,e^{- i H \big((t_i - t_j) - i (\epsilon_i- \epsilon_j)\big) }\, \mathcal{O}_j(\vec{x}_j)\, e^{- i H (t_j - i \epsilon_j)} \dots 
\end{equation}
is well defined when the Hamiltonian is bounded from below. This means that in the correlator, an operator $\mathcal{O}_i(t_i-i\epsilon_i)$ to the left of $\mathcal{O}_j(t_j-i\epsilon_j)$ requires $\epsilon_i > \epsilon_j$  to have a decaying exponential.\footnote{For a review of the $i\epsilon$ prescription for $n$-point functions and its interpretation in terms of the different Wick rotations around the branch cuts see \cite{Hartman:2015lfa}.} 

The momentum space expression for  the Euclidean 2-point function can be obtained by simply Fourier transforming the expression \eqref{2ptE}. The Euclidean 2-point function in momentum space is
\begin{equation} \label{2ptEm}
 G^{\Delta}_{\scriptscriptstyle{E}}(p) \equiv \llangle \mathcal{O}(p) \mathcal{O}(-p) \rrangle_{\scriptscriptstyle{E}} = \frac{\pi^{d/2}\, \Gamma(d/2-\Delta)}{2^{2 \Delta - d}\, \Gamma(\Delta)} \,\,|p|^{2\Delta -d},
\end{equation}
where we have removed the momentum-conserving $\delta$-function\footnote{We will often suppress the explicit dependence of $G^\Delta_{\scriptscriptstyle{E}}$ on the dimension of the operators, and write $ G_{\scriptscriptstyle{E}}(p)$. \label{dimdep}}   
\begin{equation}
 \langle \mathcal{O}(p) \mathcal{O}(q) \rangle_{\scriptscriptstyle{E}} = (2 \pi)^d \,\delta^{(d)}(p+q) \,G_{\scriptscriptstyle{E}}(p)\, ,
\end{equation}
and $d$ is the dimension of space. Despite the Fourier transform requiring $0 < \textrm{Re} \, \Delta < d/2$ for convergence, the above expression is analytic in $\Delta$ and $d$ and can therefore be analytically-continued to any conformal dimension $\Delta - d/2 \notin  \mathbb{N}^0$ above the unitarity bound $\Delta>0$.

When $\Delta - d/2 \in  \mathbb{N}^0$, the Euclidean 2-point function has no Fourier transform and must be regularised and renormalised. This renormalisation is what leads to anomalies. In position space, the 2-point function can be regularised by differential regularisation \cite{Freedman:1991tk}. The renormalised 2-point function depends on the renormalisation scale, and this dependence is given by the trace anomaly coefficient of a background scalar current \cite{OP}. In momentum space, the 2-point function can be regularised by dimensional regularisation, as is clear from the fact that the problem occurs for negative integer arguments of the $\Gamma$-function in \eqref{2ptEm}, and the renormalised 2-point function gains a logarithmic dependence in the momentum $p.$

The Lorentzian 2-point function in momentum space, in principle, follows from the Fourier transform of \eqref{2ptL}. However, this is not a straightforward calculation. It becomes much simpler though in $d=4$ because the angle integrals simplify  considerably, and we obtain\footnote{See footnote \ref{dimdep}.}
\begin{equation} \label{2ptLm-4d}
 G^{\Delta}(p) \equiv \llangle \mathcal{O}(p) \mathcal{O}(-p) \rrangle = \frac{(2\pi)^3 }{4^{\Delta - 1} \,\Gamma(\Delta -1)\, \Gamma(\Delta)} \,\theta(p^{\scriptscriptstyle{0}} - |\vec{p}\,|)\, \,| p |^{2\Delta- 4}\, ,
\end{equation}
where unambiguously
\begin{equation}
|p|=\sqrt{(p^{\scriptscriptstyle{0}})^2-|\vec{p}\,|^2}\, 
\end{equation}
thanks to the Heaviside step function,
\begin{equation}
 \theta(x) =\begin{dcases}
  \,1 \qquad  &x>0\,,\\
  \,0 \qquad & x\leq 0 \, 
 \end{dcases} \;.
\end{equation}
In all expressions where the norm of a Lorentzian momentum will appear, the argument inside the square-root can unambiguously be written without the absolute-value sign because of the presence of a step function.

The other Wightman function corresponding to the Fourier transform of $\langle \mathcal{O}(0) \mathcal{O}(x) \rangle$, i.e. the Fourier transform of \eqref{2ptL} but with negative $\epsilon$, is
\begin{align}\label{2ptLm-reverse-4d}
G^{\Delta}(-p)\equiv \llangle \mathcal{O}(-p) \mathcal{O}(p) \rrangle  =  \frac{(2\pi)^{3} }{4^{ \Delta - 1}\, \Gamma(\Delta -1) \,\Gamma(\Delta)} \, \theta(- p^{\scriptscriptstyle{0}} - |\vec{p}\,|)\, \,| p |^{2\Delta- 4}\,.
\end{align}
It becomes clear that the ordering of the operators is reflected in the sign of $p^{\scriptscriptstyle{0}}$ in the Heaviside step function. 
At a technical level, after the $\vec{x}$ integrals have been done, there is a factor of $e^{i (p^{\scriptscriptstyle{0}} - |\vec{p}\,|) t}$ for $G(p)$ or a factor of $e^{i (p^{\scriptscriptstyle{0}} + |\vec{p}\,|) t}$ for $G(-p)$, and the integrand only has a singularity in the upper or lower half $t$-plane respectively. Therefore, the integral over $t$ is zero unless $p^{\scriptscriptstyle{0}} - |\vec{p}\,| \geq 0$ or $p^{\scriptscriptstyle{0}}+  |\vec{p}\,| \leq0$ in each case, thus producing the respective step functions.

It is straightforward to see that since the operators are the same, swapping the ordering just amounts to swapping their respective momenta, $p$ and $-p$.

These $d=4$ Wightman functions already make it clear that the Wick rotation from Euclidean to Lorentzian signature in momentum space is not as straightforward as the $i \epsilon$ prescription in position space. Indeed, even if one can `prescribe'  the correct step functions (as following from the above singularity-based or support arguments), the $\Delta$- and $d$-dependent coefficients are hard to predict. These are bound to be even more complicated for general $d$, where the Fourier transform is much harder to compute. Furthermore, for higher-point functions even `prescribing' the correct step functions becomes difficult.
Therefore, a pertinent question is how to obtain the Wightman 2-point functions, without directly Fourier-transforming,  from the Euclidean 2-point function \eqref{2ptEm}. We address this next.

Consider the Euclidean 2-point function as a Fourier transform,
\begin{equation} \label{FT2pt}
G_{\scriptscriptstyle{E}}(x) = \int\limits \frac{d^d p}{(2 \pi)^d}\, e^{i p \cdot x}\, G_{\scriptscriptstyle{E}}(p)\,,
\end{equation}
where it is understood that the volume element and the inner product in the exponential are Euclidean.
As reviewed above, the Lorentzian function in position space is easily obtained by a Wick rotation together with the $i\epsilon$ prescription, hence we can write:
\begin{equation} \label{FT2pt2}
G(x)
= G_{\scriptscriptstyle{E}}(i (t- i \epsilon), \vec{x}) =  \int \frac{d^{d-1} \vec{p}}{(2 \pi)^{d-1}}  \,e^{i \vec{p} \cdot \vec{x}} \int\limits_{-\infty}^\infty \frac{d p_{\scriptscriptstyle{E}}}{2 \pi} \,e^{ -p_{\scriptscriptstyle{E}} (t- i \epsilon)} \,G_{\scriptscriptstyle{E}}(p_{\scriptscriptstyle{E}}, \vec{p}\,)\,,
\end{equation}
where for the last equality we have used  \eqref{FT2pt}. 

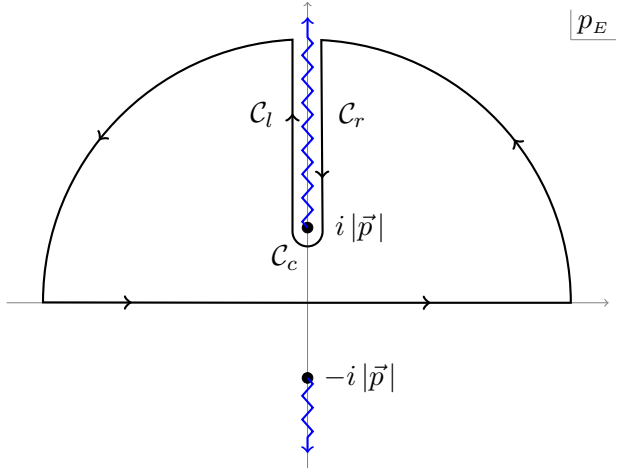
\begin{figure}
\centering
\begin{tikzpicture}[decoration={markings,
mark=at position 0.07 with {\arrow[line width=1pt]{>}},
mark=at position 0.25 with {\arrow[line width=1pt]{>}},
mark=at position .456 with {\arrow[line width=1pt]{>}},
mark=at position 0.58 with {\arrow[line width=1pt]{>}},
mark=at position 0.75 with {\arrow[line width=1pt]{>}},
mark=at position 0.9 with {\arrow[line width=1pt]{>}}
}
]
\draw[help lines,->] (-4,0) -- (4,0) coordinate (xaxis);
\draw[help lines,->] (0,-2.2) -- (0,4) coordinate (yaxis);
\draw[fill] (0,1) circle (2pt);
\draw[fill] (0,-1) circle (2pt);
\draw[color=blue,line width=0.8pt,->,decorate,decoration={zigzag,amplitude=2,segment length=9,post length=4}] (0,1) -- (0,3.8);
\draw[color=blue,line width=0.8pt,->,decorate,decoration={zigzag,amplitude=2,segment length=9,post length=4}] (0,-1) -- (0,-2);
\draw[line width=0.8pt,postaction=decorate] (0,0)  -- (3.5,0) arc (0:87:3.5) -- (0.2,0.95) arc (0:-180:0.2)  -- (-0.2,3.5) arc (93:180:3.5) -- (0,0);
\draw[help lines] (3.5,3.5) -- (3.5,3.9);
\draw[help lines] (3.5,3.5) -- (4.1,3.5);
\node at (0.7,1) {$i\, |\vec{p}\,|$};
\node at (0.7,-1) {$-i\,|\vec{p}\,|$};
\node at (0.6,2.5) {$\C_r$};
\node at (-0.6,2.5) {$\C_l$};
\node at (-0.3,0.6) {$\C_{c}$};
\node at (3.8,3.7) {$p_{\scriptscriptstyle{E}}$};
\end{tikzpicture}
\caption{Closed contour of integration. The contribution from the arc at infinity vanishes.} \label{fig:contour-2pnt}
\end{figure}

The integral above is no longer a Fourier transform, but it can be recast into a Fourier transform by Wick-rotating $p_{\scriptscriptstyle{E}}$ to the imaginary axis. However, this must be done carefully, taking into consideration the analytical properties of $G_{\scriptscriptstyle{E}}(p_{\scriptscriptstyle{E}},\vec{p}\,)$ on the complex $p_{\scriptscriptstyle{E}}$ plane. Indeed from \eqref{2ptEm} it is clear that for general $\Delta$, $G_{\scriptscriptstyle{E}}(p_{\scriptscriptstyle{E}}, \vec{p}\,)$ has branch points at $p_{\scriptscriptstyle{E}} = \pm i |\vec{p}\,|$ 
and $\infty.$ 
We consider then extending the contour of integration along the real axis, into the upper half of the complex plane and around the branch cut from $i |\vec{p}\,|$ to $i \infty$, forming the closed contour depicted in figure \ref{fig:contour-2pnt}. Because of the $i\epsilon$ term in the exponential with $\epsilon>0$, the integral over the arc at infinity vanishes. If the operators were ordered oppositely so that $\epsilon$ were negative, we would instead close the contour on the lower half complex plane to get a vanishing contribution from the arc at infinity.  

Since the function is analytic inside the closed contour, the integral in \eqref{FT2pt2} is equal to that along the contours on each side of the branch cut ($\C_r$ and $\C_l$ in figure \ref{fig:contour-2pnt}) and the cup $\C_c$ around the branch point, hence
\begin{equation}\label{eqn_random}
G(x) =  -\int \frac{d^{d-1} \vec{p}}{(2 \pi)^{d-1}}\,e^{i \vec{p} \cdot \vec{x}} \,
	 \int\limits_{\C} \frac{d p_{\scriptscriptstyle{E}}}{2 \pi} \,e^{ -p_{\scriptscriptstyle{E}} (t- i \epsilon)} \, G_{\scriptscriptstyle{E}}(p_{\scriptscriptstyle{E}}, \vec{p}\,)\,, \qquad\, \C=\C_r\,\cup\,\C_l\, \cup\, \C_c .
\end{equation}
Changing now the integration variable to  $p^{\scriptscriptstyle{0}}=-i p^{\scriptscriptstyle{E}}$ implements the Wick rotation for the time component of the momentum. Taking into account the phase difference on both sides of the cut, the $p_{\scriptscriptstyle{E}}$ integral becomes 
\begin{align} \label{interm}
	 - 2\, \sin\left(\pi (\Delta - d/2)\right) \frac{\pi^{d/2} \,\Gamma(d/2-\Delta)}{2^{2 \Delta - d}\, \Gamma(\Delta)} 
	 \int\limits_{|\vec{p}\,|}^{\infty} dp^{\scriptscriptstyle{0}} \,e^{ -i p^{\scriptscriptstyle{0}} (t- i \epsilon)} \,  \left( (p^{\scriptscriptstyle{0}})^2 - |\vec{p}\,|^2 \right)^{\Delta- d/2} \, .
\end{align}
For $\Delta > d/2-1$, the contribution from the cup $\C_c$ vanishes and the $p_{\scriptscriptstyle{E}}$ integral in \eqref{eqn_random} is equivalent to expression \eqref{interm}. However, for  $\Delta \leq d/2-1$, the above integral is divergent and rendered finite by contributions from the integral along contour $\C_c$. Our notation in \eqref{interm} is such that it includes contributions that render the integral finite.

Plugging expression \eqref{interm}  back into \eqref{eqn_random} gives $G(x)$ as a Fourier transform
\begin{equation}
G(x) = \int \frac{d^{d} p}{(2 \pi)^{d}} \, e^{i p \cdot x} \, \frac{\pi^{d/2+1} \, \theta(p^{\scriptscriptstyle{0}} - |\vec{p}\,|)}{2^{2 \Delta - d-1}\, \Gamma(\Delta -d/2+1) \,\Gamma(\Delta)} \,\,|p|^{2\Delta- d}\, . \label{2ptLp}
\end{equation}
From the integrand we can read off the  Wightman 2-point function in momentum space 
\begin{align}\label{2ptLm}
G^{\Delta}(p)\equiv \llangle \mathcal{O}(p) \mathcal{O}(-p) \rrangle  =  \frac{\pi^{d/2+1} }{2^{2 \Delta - d-1}\, \Gamma(\Delta -d/2+1) \,\Gamma(\Delta)} \, \theta(p^{\scriptscriptstyle{0}} - |\vec{p}\,|)\, \,| p |^{2\Delta- d}\,.
\end{align}
The above reproduces expression \eqref{2ptLm-4d}  in $d=4$ dimensions obtained by Fourier transforming.

The other Wightman function can be found by the complex conjugation of \eqref{2ptLp} and letting $p \rightarrow -p$ in the integral, since it corresponds to letting $i\epsilon\rightarrow-i\epsilon$ on the right-hand side of \eqref{2ptL}, and reads
 \begin{align}\label{2ptLm-reverse}
G^{\Delta}(-p)\equiv \llangle \mathcal{O}(-p) \mathcal{O}(p) \rrangle  =  \frac{\pi^{d/2+1} }{2^{2 \Delta - d-1}\, \Gamma(\Delta -d/2+1) \,\Gamma(\Delta)} \, \theta(-p^{\scriptscriptstyle{0}} - |\vec{p}\,|)\, \,| p |^{2\Delta- d}\,.
\end{align}
Both Wightman functions vanish when the momentum is spacelike or null, so they only have support in the  future momentum-space light-cone in the case of $G(p)$,  or the past momentum-space light-cone in the case of $G(-p)$. 

As opposed to the Euclidean 2-point function \eqref{2ptEm}, which has a $\Gamma$-function in the numerator, a factor of $\Gamma(\Delta-d/2+1)$ appears now in the denominator of the Lorentzian 2-point function, \eqref{2ptLm}. Therefore it does not diverge for particular values of $\Delta$, and hence does not require renormalisation.

With this second method of computing the momentum-space 2-point function by means of a formal Wick rotation, the Heaviside step function appears as a result of deforming the contour around the branch point in the complex plane where the momentum becomes null. This is analogous to the fact that in position space the ordering of operators in Lorentzian correlators is given by how one analytically continues around branch points corresponding to points where operators become null-separated, and hence no longer commute with each other by microcausality \cite{Hartman:2015lfa, Haag:1992hx}.

This second method has the advantage that it does not require us to perform complicated integrals, as opposed to the Fourier transform method. 

In deriving \eqref{2ptLm}, we have assumed that $\Delta>0$ and $\Delta- d/2$  is non-integer. The positivity condition is required because the Euclidean Fourier transform \eqref{2ptEm} from which we start our analysis, is derived using Schwinger parametrisation \eqref{schwinger}. However, we observe that both the Euclidean, \eqref{2ptEm}, and the Lorentzian, \eqref{2ptLm},  2-point functions can be analytically continued to negative $\Delta$ as long as it is not an integer.
The condition that $\Delta>0$ is a natural consequence of unitarity, but as we will see in the next section, in order to derive 3-point functions in momentum space for particular values of the conformal dimensions, we require the expression for  $G$ for $\Delta<0$. Therefore, we need to consider the case when $- \Delta \in \mathbb{N}^0$. 

The second condition, $\Delta- d/2$ a non-integer, is required because in performing the Wick rotation, we have assumed the existence of branch points, while an integer exponent $\Delta- d/2$ leads to poles instead. 
When $\Delta- d/2 \in \mathbb{N}^0$, we have already remarked that the Euclidean expression requires regularisation and renormalisation, hence in this case, the formal Wick rotation method cannot be implemented as above. However, the Lorentzian expression \eqref{2ptLm} as computed by Fourier transform is nevertheless valid for these cases, essentially because the $i \epsilon$ prescription precludes the existence of contact singularities in the correlator. 
Therefore, we only consider separately the case $d/2 - \Delta \in \mathbb{N}$. 

We consider these two cases below.

\vspace{3mm}

{\underline{\bf{Case I: $- \Delta  \in \mathbb{N}^0$}}}

\vspace{3mm}

Renaming $- \Delta =n \in \mathbb{N}^0$, the Euclidean 2-point function can simply be found in this case by using
\begin{equation}
 \int d^d x\, x^{2 n}\, e^{- i p \cdot x}  = (2 \pi)^d \left(-\partial^{2}_{p}\right)^n \delta^{(d)}(p)\,, \quad  \; n \in \mathbb{N}^0. \label{identity1}
\end{equation}
In Lorentzian signature we have the same expression with $x^2$ being the Lorentzian norm, and hence with 
\begin{equation}
 \partial^{2}_{p} = \eta^{\mu \nu} \partial_{p^\mu} \partial_{p^\nu}\, .
\end{equation}

{\underline{\bf{Case II: $d/2 - \Delta \in \mathbb{N}$}}} 

\vspace{3mm}

In this case, there are two possibilities. If $d$ is an even integer, for large enough $\mathbb{N}$ such that $\Delta<0$, this case becomes the same as case I. For small enough $\mathbb{N}$ such that $\Delta>0$, or more generically when $d$ is not an even integer, this case is different from the previous, and we address it in the following. 

Since the Euclidean 2-point function \eqref{2ptEm} is well-defined when $d/2 - \Delta = n \in \mathbb{N}$, we can proceed as in the general case and perform a Wick rotation so as to rewrite \eqref{FT2pt2} as a Lorentzian Fourier transform. 

The $p_{\scriptscriptstyle{E}}$ integral  in (c.f.\ \eqref{FT2pt2})
\begin{equation} 
G(x)
=   \frac{2^{2n} \,\pi^{d/2}\, \Gamma(n)}{ \Gamma(d/2-n)}  \int \frac{d^{d-1} \vec{p}}{(2 \pi)^{d-1}}  e^{i \vec{p} \cdot \vec{x}}\int\limits_{-\infty}^\infty \frac{d p_{\scriptscriptstyle{E}}}{2 \pi} \, \frac{e^{ -p_{\scriptscriptstyle{E}} (t- i \epsilon)} }{ |p|^{2 n}}\,,
\end{equation}
can now no longer be rotated in the same way, since 
its integrand has poles rather than branch cuts.
However, this integral is now easier as it can be evaluated by the residue theorem. The resulting expression can then be rewritten in terms of an integral over $p^{\scriptscriptstyle{0}}$ with a Dirac $\delta$-function, and gives
\begin{equation}\label{k-integral-E-sp}
\int\limits_{-\infty}^\infty 
\frac{dp_{\scriptscriptstyle{E}}}{2\pi}\,\frac{e^{-p_{\scriptscriptstyle{E}}\,(t-i\ep)}}{(p_{\scriptscriptstyle{E}}^2+|\vec{p}\,|^2)^{n}}
=\frac{2\pi}{\Gamma(n)}\int\limits_{-\infty}^\infty \frac{dp^{\scriptscriptstyle{0}}}{2 \pi}\,\frac{e^{-i p^{\scriptscriptstyle{0}}\,(t-i\ep)}}{(p^{\scriptscriptstyle{0}}+|\vec{p}\,|)^n}\,\partial_{p^{\scriptscriptstyle{0}}}^{n-1}\delta(p^{\scriptscriptstyle{0}}-|\vec{p}\,|)\, .
\end{equation}
In the above step, the derivatives acting on the Dirac $\delta$-function come from integrating by parts the derivatives coming from the residue of the integral. Therefore, for $d/2 - \Delta = n \in \mathbb{N}$
\begin{equation} \label{k-integral-sp}
G(p)  =  \frac{2^{2n+1} \,\pi^{d/2+1}}{ \Gamma(d/2-n)}   \,
\frac{\partial_{p^{\scriptscriptstyle{0}}}^{n-1}\delta(p^{\scriptscriptstyle{0}}-|\vec{p}\,|)}{(p^{\scriptscriptstyle{0}}+|\vec{p}\,|)^n}\,.
\end{equation}

\vspace{3mm}

In summary, from equations \eqref{2ptLm}, \eqref{identity1} and \eqref{k-integral-sp}, the Lorentzian 2-point function of scalars with dimensions $\Delta$ is
\begin{equation} \label{FT2ptg}
G^\Delta(p) = \int d^{d} x \, \frac{e^{-i p \cdot x} }{\big({- (t - i \, \epsilon)^2 + | \vec{x}|^2}\big)^{\Delta}}\,  =  \begin{dcases}
            (2 \pi)^d \left(-\partial^{2}_{p}\right)^{-\Delta} \delta^{(d)}(p) & -\Delta \in \mathbb{N}^0  \\[3pt]                                                                                             
    \frac{\pi^{d/2+1}}{2^ {2 \,\Delta-d-1} \, \Gamma(\Delta)}   \,
\frac{\partial_{p^{\scriptscriptstyle{0}}}^{d/2 - \Delta-1}\delta(p^{\scriptscriptstyle{0}}-|\vec{p}\,|)}{(p^{\scriptscriptstyle{0}}+|\vec{p}\,|)^{d/2 - \Delta}} & d/2 - \Delta \in \mathbb{N} \\[3pt] 
\frac{\pi^{d/2+1} \, \theta(p^{\scriptscriptstyle{0}} - |\vec{p}\,|)}{2^{2 \Delta - d-1}\, \Gamma(\Delta -d/2+1) \,\Gamma(\Delta)} \,\,|p|^{2\Delta- d} & \textrm{otherwise}.
\end{dcases}
\end{equation}

\section{3-point function of scalars}
\label{sec:sca}

In this section we compute the Lorentzian 3-point function of scalar operators in momentum space. We obtain two different expressions for it, each of which we present in the next two subsections: the momentum-integrated expression, which is an integrated product of three 2-point functions, and the triple-Bessel expression, which follows from the Wick rotation of the analogous triple-K Euclidean correlator. In appendix \ref{sec:scalar-FT} we compute the 3-point correlator from a direct Fourier transform, albeit for the case of $d=4$. 

\subsection{Momentum-integrated expression} \label{sec:scam}

The 3-point function of scalar operators in position space is given by~\cite{OP},
\begin{equation}
\langle \mathcal{O}_1(x_1)\,\mathcal{O}_2(x_2)\,\mathcal{O}_3(x_3)\rangle
= \frac{c_{123}}{(x_{23}^2)^{\beta_1}\,(x_{13}^2)^{\beta_2}\,(x_{12}^2)^{\beta_3}}\, ,
\end{equation}
where
\begin{gather}\label{}
 \beta_j=\frac{\Delta_t}{2}- \Delta_j\,, \qquad \D_t=\D_1+\D_2+\D_3\,, \qquad  \Delta_t=2\,\beta_t\,.
\end{gather}
The conformal dimension of operator $\mathcal{O}_j$ is $\Delta_j$ and the norm $|x_{ij}|^2$ is the Euclidean, Lorentzian norm of the separation $x_{i}-x_{j}$ in Euclidean, Lorentzian signature respectively. In Lorentzian signature we also specify an $i \epsilon$ prescription, $t_j-i\epsilon_j$, that dictates the ordering of operators as discussed in the previous section. In the above correlator it is $\epsilon_1>\epsilon_2>\epsilon_3$.
The factor $c_{123}$ is the structure constant and depends on the conformal dimensions $\D_j$ of the operators.

The Fourier transform, and hence the correlator in momentum space is then given by 
\begin{align}
\langle \mathcal{O}_1(p_1)\,\mathcal{O}_2(p_2)\,\mathcal{O}_3(p_3)\rangle
&= \int \prod_{j=1}^3 \left( d^dx_j \,  e^{- i p_j \cdot x_j} \right) \frac{c_{123} }{(x_{23}^2)^{\beta_1}\,(x_{13}^2)^{\beta_2}\,(x_{12}^2)^{\beta_3}}\, \notag \\[7pt]
&= (2 \pi)^d \, \delta^{(d)}(p_1 + p_2 +p_3) \,  C(p_1, p_2; \{\beta_j\})\,,  \label{mcorreldef}
\end{align}
where
\begin{equation}\label{delta-fn-stripped}
  C(p_1, p_2; \{\beta_j\})  =  c_{123}\int d^d x_1  d^dx_2 \,    \frac{ e^{- i p_1 \cdot x_1} e^{- i p_2 \cdot x_2}}{(x_{2}^2)^{\beta_1}\,(x_{1}^2)^{\beta_2}\,(x_{12}^2)^{\beta_3}}\,.
\end{equation}
In the above,  we have used the translation invariance of the position space correlator to extract a momentum preserving $\delta$-function. The $\delta$-function stripped correlator is also denoted using double angle brackets in the literature,
\begin{equation}
 C(p_1, p_2; \{\beta_j\}) \equiv \llangle \mathcal{O}_1(p_1)\,\mathcal{O}_2(p_2)\,\mathcal{O}_3(p_3)\rrangle\,.
\end{equation}
We will suppress the momentum dependence of $C(p_1, p_2; \{\beta_j\})$, \emph{viz}.\ write $C(\{\beta_j\})$, when the momentum dependence is clear.

At the expense of introducing an extra momentum integral, we can reexpress the $\delta$-function stripped correlator \eqref{delta-fn-stripped} as a product of three 2-point functions
\begin{equation} \label{3pt}
 C(\{\beta_j\}) =  c_{123} \int \frac{d^d k}{(2 \pi)^d}\,  G^{\beta_1}(p_2 +k) \, G^{\beta_2}(p_1-k) \, G^{\beta_3}(k)\,.
\end{equation}
This is very convenient because we can now determine the 3-point correlator from the expressions of the 2-point function presented in section \ref{sec:pre}.

The same holds in the Euclidean case for $C_{\scriptscriptstyle{E}}(\{\beta_j\})$, but with the 2-point functions above given by $G_{\scriptscriptstyle{E}}^{\beta_j}$. In this case, when the $\beta_j$ are general, we insert the 2-point function result \eqref{2ptEm}, and the resulting $\delta$-function stripped correlator is
\begin{equation}
C_{\scriptscriptstyle{E}}(\{\beta_j\})
=c_{123}\, \pi^{3d/2}\,2^{3d-2\beta_t}\,\prod\limits_{j=1}^3\frac{\Gamma(d/2 - \beta_j)}{\Gamma(\beta_j)}\,\int \frac{d^dk}{(2\pi)^d}\,
\frac{1}{|p_2+k|^{d-2 \beta_1}\,|p_1-k|^{d- 2\beta_2}\,|k|^{d- 2 \beta_3}}\, . \label{E3pt}
\end{equation}
This expression for the Euclidean 3-point function exhibits divergences in certain cases, either in the $\Gamma$-functions in the (numerator or denominator of the) coefficient, and/or in the momentum integral, and hence requires regularisation. Some of the singularities cancel with each other, rendering the above expression finite, others simply remain and require renormalisation in order to obtain a finite 3-point function. The analysis of the cases (sets of $\beta_j$) which require renormalisation has been carried out in~\cite{BMS:renoms}, and it becomes easier to do in terms of the triple-K expression for $C_{\scriptscriptstyle{E}}(\{\beta_j\})$. We therefore postpone further comments to the next subsection.

\vspace{5mm}

We go back to the Lorentzian case. Since the 3-point function \eqref{3pt} is given in terms of the 2-point functions $G^{\beta_j}$, we get different expressions depending on the value of $\beta_j$. We first consider the case where the $\beta_j$ are general, separately from those when they satisfy any of the special cases I and II discussed in section \ref{sec:pre}, for which the 2-point function is given in terms of a Dirac $\delta$-function. 

\vspace{3mm}

\noindent
{\underline{\bf{General case:  $-\beta_j  \not\in\{  \mathbb{N}-\frac{d}{2}\,,\,\mathbb{N}^0\}$}}}

\vspace{3mm}

Using equations \eqref{3pt} and \eqref{2ptLm},
\begin{align}
 C(p_1, p_2; \{\beta_j\})=
c(\{\beta_j\})
& \int \frac{d^dk}{(2\pi)^d}\,
\frac{\theta(p_{2}^{\scriptscriptstyle{0}}+k^{\scriptscriptstyle{0}}-|\vec{p}_2+\vec{k}|)\,\theta(p_1^{\scriptscriptstyle{0}}-k^{\scriptscriptstyle{0}}-|\vec{p}_1-\vec{k}|)\,\theta(k^{\scriptscriptstyle{0}}-|\vec{k}|)}{|p_2+k|^{d - 2 \beta_1}\,|p_1-k|^{d - 2 \beta_2}\,|k|^{d - 2 \beta_3}}\,  \label{eq1}
\end{align}
with
\begin{equation}
  c(\{\beta_j\}) = c_{123}\,
 \frac{2^{3(d+1)-2\beta_t} \, \pi^{3(d/2+1)} }{\prod\limits_{j=1}^3\Gamma( \beta_j- d/2 +1)\, \Gamma(\beta_j)}\, . \label{cdef}
\end{equation}
This is the generic Lorentzian 3-point function. Notice that  the step functions in the integrand imply  an overall
\begin{equation}
\theta(p_1^{\scriptscriptstyle{0}}-|\vec{p}_1|)\theta(-p_3^{\scriptscriptstyle{0}}-|\vec{p}_3|)\, .
\end{equation}
In other words, as for the 2-point function, the correlator vanishes unless the left-most operator has a positive energy and the right-most operator has a negative energy.
In particular, the 3-point correlator vanishes when $p_1=0$ or $p_3=0$. For $\beta_2+\beta_3>1/2,$ this can be understood from the formal expression of the Fourier transform  \eqref{mcorreldef}. When $p_1=0$, the transform involves the integral
\begin{equation}
\langle \mathcal{O}_1(0)\,\mathcal{O}_2(p_2)\,\mathcal{O}_3(p_3)\rangle
\sim\int d^dx_1\,\frac{1}{(x_{13}^2)^{\beta_2}\,\,(x_{12}^2)^{ \beta_3}}\, .\nonumber
\end{equation}
Since $\epsilon_1>\epsilon_2>\epsilon_3$, the $t_1$ integrand has four branch points in the upper half plane and is analytic in the lower half. The integral contour can then be closed on the lower half plane, which shows that the above integral vanishes. Similarly, if $p_3=0$, the branch points all lie in the lower half plane, and enclosing the contour on the upper half plane gives a vanishing integral. In the case $p_2=0$, two branch points lie on the upper half plane, and two in the lower, therefore it does not immediately follow that the integral vanishes.

When $-  \beta_j \in \mathbb{N}^0$ or $ d/2 - \beta_j\in \mathbb{N}$, the 2-point functions are no longer given by \eqref{2ptLm} but rather by the first two cases in \eqref{FT2ptg}---the special cases I and II of the 2-point function studied in section \ref{sec:pre}. We consider these cases next.

\vspace{3mm}

\noindent
{\underline{\bf{Case I: $-  \beta_j \in \mathbb{N}^0$}}}

\vspace{3mm}

Notice that this can only be the case for one of the $\beta_j$. For two or all of the $\beta_j$ to be simultaneously non-positive, at least one of the conformal dimensions $\Delta_j$ would have to be negative, and this would violate the unitarity bound. Without loss of generality we let $\beta_1 = - n$, with $n \in \mathbb{N}^0.$ 
In this case, using \eqref{3pt} and the first line of \eqref{FT2ptg}, we find
\begin{equation}
 C(-n, \beta_2, \beta_3)=  c(-n, \beta_2, \beta_3) \left(-\partial^{2}_{p_2}\right)^n \,  \frac{\theta(p_{1}^{\scriptscriptstyle{0}}+p_2^{\scriptscriptstyle{0}}-|\vec{p}_1+\vec{p}_2|)\,\theta(- p_2^{\scriptscriptstyle{0}}-|\vec{p}_2|)}{|p_1+p_2|^{d- 2 \, \beta_2}\, |p_2|^{d- 2 \, \beta_3}} \,, \label{L-int}
\end{equation}
where
\begin{equation}
c(-n, \beta_2, \beta_3) =c_{123}\,
 \frac{2^{2(d- \beta_2 - \beta_3+1)} \, \pi^{d+2} }{\prod\limits_{j=2}^3 \Gamma(\beta_j- d/2 + 1)\,\Gamma(\beta_j)}\, .
\end{equation}

\noindent
{\underline{\bf{Case II: $ d/2 - \beta_j\in \mathbb{N}$}}}

\vspace{3mm}

If $d$ is an even integer, for large enough $\mathbb{N}$ such that $\beta_j<0$, this case becomes the same as the previous one. For small enough $\mathbb{N}$ such that $\beta_j>0$, or more generically when $d$ is not an even integer, this case is different from the previous, and the expression for the 2-point function to be used is given by \eqref{k-integral-sp}.

Without loss of generality we let  $\beta_1=d/2 -n$, with $n\in\mathbb{N}$. Using equation \eqref{3pt} and \eqref{k-integral-sp},  
\begin{align}
 C(d/2-n, \beta_2, \beta_3)= c(d/2-n, \beta_2, \beta_3)
  \int \frac{d^dk}{(2\pi)^d}\,&
  \frac{\theta(p_1^{\scriptscriptstyle{0}}-k^{\scriptscriptstyle{0}}-|\vec{p}_1-\vec{k}|)\,\theta(k^{\scriptscriptstyle{0}}-|\vec{k}|)}{\,|p_1-k|^{d - 2 \beta_2}\,|k|^{d - 2 \beta_3}} \nonumber\\[3pt]
& \times \frac{\partial_{p_{2}^{\scriptscriptstyle{0}}}^{n-1}\delta(p_{2}^{\scriptscriptstyle{0}}+k^{\scriptscriptstyle{0}}-|\vec{p}_2+\vec{k}|)}{(p_{2}^{\scriptscriptstyle{0}}+k^{\scriptscriptstyle{0}}+|\vec{p}_2+\vec{k}|)^n}\,, \label{eq1-sp}
\end{align}
with
\begin{equation}
c(d/2-n, \beta_2, \beta_3)=c_{123}\,
 \frac{4^{d+n- \beta_2 - \beta_3+3/2} \, \pi^{3(d/2+1)} }{\Gamma(d/2-n)\prod\limits_{j=2}^3 \Gamma( \beta_j- d/2 +1)\,\Gamma(\beta_j)}\, .
\end{equation}

Since $\beta_j = d/2-n$ with $n$ a positive integer can be positive for small enough $n$, it is possible for two or all of the $\{\beta_j\}$ to fall into case II.  It is also possible to have one of the $\{\beta_j\}$ satisfy case I. The expression in these cases follows likewise from the appropriate replacements of the 2-point function \eqref{FT2ptg} in expression \eqref{3pt}.

\vspace{5mm}

If we insisted on using the general expression \eqref{eq1} for the special $\beta_j$ cases, we would find that the coefficient $c(\{\beta_j\})$ in \eqref{cdef} vanishes due to $\Gamma$-functions with negative integer arguments in the denominator, and that the momentum integral diverges. An appropriate regularisation would then be required to obtain a finite expression for the 3-point function.  Considering instead the appropriate form of the 2-point functions that go into \eqref{3pt} ensures that the obtained expression for the 3-point function has a non-vanishing coefficient. In principle, the momentum integral of the final expression could still diverge, but it is in fact always finite (possibly obtained by analytical continuation) because, as we will show in the next subsection, the 3-point function is finite without requiring renormalisation.

\subsection{Triple-Bessel expression} \label{sec:tbe}

Equation \eqref{3pt} shows that the 3-point function, both in Euclidean and Lorentzian signatures, can be written as the integral over an auxiliary momentum of a triple product of 2-point functions. In Euclidean space, the 3-point function can also be written as a single integral over a triple product of modified Bessel functions of the second kind $K_\nu$ \cite{Barnes:2010jp,BMS:imp},
\begin{equation}
C_{\scriptscriptstyle{E}}(p_1, p_2 ;\{\beta_j\})
= c_{\scriptscriptstyle{E}}(\{\beta_j\}) \,\int\limits_0^\infty dt \, t^{d/2-1}  \, \prod_{j=1}^{3}\, |p_j|^{\nu_j}\,K_{\nu_j}(|p_j| \, t) \,, \label{3KE}
\end{equation}
where
\begin{equation}
 c_{\scriptscriptstyle{E}}(\{\beta_j\}) = c_{123}\,\frac{ 2^{4+3d/2-\Delta_t}\,\pi^{d}}{\Gamma(\frac{\Delta_t - d}{2})\,\prod\limits_{j=1}^3\Gamma(\beta_j)}, \qquad \nu_j = \Delta_j- \frac{d}{2} = \frac{\Delta_t- d}{2} - \beta_j  \label{cEdef}
\end{equation}
and $p_3 = - (p_1+ p_2).$  

This expression for the Euclidean 3-point function is divergent for certain sets of the parameters $\beta_j$, and requires regularisation and in some cases further renormalisation. An advantage of this expression over the momentum-integrated one \eqref{E3pt} is that all $\Gamma$-functions in the coefficient appear now in the denominator. This means that the divergences can only come from the $t$-integral, which simplifies the analysis.  
The $t$-integral diverges when
\begin{equation}\label{conv-condE}
d/2< \sum_{j=1}^3 |\nu_j|.
\end{equation}
This is due to the behaviour of the integrand at small $t$, but in general the integral can be defined by analytical continuation. However, when 
\begin{equation}\label{div-condition}
\frac{d}{2}\pm \nu_1\pm\nu_2\pm\nu_3 =-2n, \qquad n\in \mathbb{N}^0
\end{equation}
is satisfied for any independent choice of the $\pm$ signs in the three terms, the analytical continuation fails \cite{BMS:imp}.

For the cases which satisfy condition \eqref{div-condition} with the sign choices $(---)$ and $(+--)$ (and permutations), the divergence of the integral implies that the above expression for $C_{\scriptscriptstyle{E}}(\{\beta_j\})$ also diverges and hence requires renormalisation; a thorough analysis has been performed in~\cite{BMS:renoms}. However, in the cases which satisfy the other two sign choices, the divergence turns out to cancel  with the divergence in the denominator of the coefficient. Indeed, the $(+++)$ choice implies $\Delta_t=d-2n$, in which case the first $\Gamma$-function in the denominator diverges. The $(-++)$ choice (or permutations) implies $\beta_1=-n$, in which case $\Gamma(\beta_1)$ in the denominator also diverges. So for the $\{\beta_j\}$ cases which satisfy \eqref{div-condition} with at least two $+$ signs, $C_{\scriptscriptstyle{E}}(\{\beta_j\})$ is finite despite the divergence of the $t$-integral, and is given by regularisation.

\vspace{5mm}

We now turn to the Lorentzian case. The method used in the Euclidean case to arrive at the triple-$K$ formula \eqref{3KE} from the momentum integrated one \eqref{E3pt} fails in the Lorentzian case essentially because $p^2_j$ is no longer positive definite. 
Nevertheless, we can again compute the analogous expression for the Lorentzian 3-point function by means of a formal Wick rotation of the Euclidean triple-K expression. As we emphasised in section \ref{sec:pre}, in order to do this we must take into account the analytic properties of \eqref{3KE}. 

The starting point is
\begin{align}\label{FT-EL-3}
\langle \mathcal{O}_1(x_1)\,\mathcal{O}_2(x_2)\,\mathcal{O}_3(x_3)\rangle
&=\langle \mathcal{O}_1(t^{\scriptscriptstyle{E}}_1,\vec{x}_1)\,\mathcal{O}_2(t^{\scriptscriptstyle{E}}_2,\vec{x}_2)\,\mathcal{O}_3(t^{\scriptscriptstyle{E}}_3,\vec{x}_3)\rangle_{\scriptscriptstyle{E}}\,\biggr|_{t^{\scriptscriptstyle{E}}_j=i\,(t_j-i\,\epsilon_j)}\nonumber\\
&=\int \frac{d^dp_1}{(2\pi)^d}\frac{d^dp_2}{(2\pi)^d}\,\,C_{\scriptscriptstyle{E}}(p_1, p_2 ;\{\beta_j\})\, e^{i\,p_1\cdot x_{13}}\,e^{i\,p_2\cdot x_{23}}\,\biggr|_{t^{\scriptscriptstyle{E}}_j=i\,(t_j-i \, \epsilon_j)}\;.
\end{align}
In the first line, the Lorentzian correlator is written in terms of the Euclidean one but with Wick-rotated time coordinates and with the appropriate $i\epsilon$ prescription, $\epsilon_1>\epsilon_2>\epsilon_3$. In the second line, the Euclidean 3-point function has been rewritten in terms of its Fourier transform, therefore the momenta are $p_j=(p_j^{\scriptscriptstyle{E}},\vec{p}_j)$, and the inner product is $p\cdot x=p^{\scriptscriptstyle{E}} \,\tau+\vec{p}\cdot\vec{x}$.

Inserting the Wick-rotated time coordinates,\footnote{We denote the 3-point function \eqref{3KE} as $C_{\scriptscriptstyle{E}}(p_1,p_2),$ notationally dropping the explicit dependence on the conformal dimensions of the operators.}
\begin{align}\label{FT-EL-3b}
\langle \mathcal{O}_1(x_1)\,\mathcal{O}_2(x_2)\,\mathcal{O}_3(x_3)\rangle
=\int \frac{d^dp_1}{(2\pi)^d}\frac{d^dp_2}{(2\pi)^d}\,C_{\scriptscriptstyle{E}}(p_1, p_2)\, e^{-\,p^{\scriptscriptstyle{E}}_1 (t_{13} - i \, \epsilon_{13})}\,e^{-\,p^{\scriptscriptstyle{E}}_2 (t_{23} - i \,\epsilon_{23})}\, e^{i\,\vec{p}_1\cdot \vec{x}_{13}}\,e^{i\,\vec{p}_2\cdot \vec{x}_{23}}\,.
\end{align}
The Lorentzian correlator in momentum space is now  found by rewriting the above integrals as a (Lorentzian) Fourier transform, i.e.\ we need to Wick rotate $p^{\scriptscriptstyle{E}}= i \, p^{\scriptscriptstyle{0}}.$

We assume that the CFT is interacting and hence the operator dimensions are non-integer and the $\nu_j$ are generic.  
We consider the integral over $p^{\scriptscriptstyle{E}}_1$ first. From \eqref{3KE},  $C_{\scriptscriptstyle{E}}(p_1, p_2)$ has branch points on the $p^{\scriptscriptstyle{E}}_1$ complex plane at
\begin{equation}
 p^{\scriptscriptstyle{E}}_1 = \pm \, i \, |\vec{p}_1|, \qquad p^{\scriptscriptstyle{E}}_1 = -p^{\scriptscriptstyle{E}}_2 \pm \,  i \,  |\vec{p}_1 + \vec{p}_2|\, ,
\end{equation}
and at $ \infty$.

\begin{figure}
\centering
\begin{tikzpicture}[decoration={markings,
mark=at position 0.07 with {\arrow[line width=1pt]{>}},
mark=at position 0.25 with {\arrow[line width=1pt]{>}},
mark=at position .456 with {\arrow[line width=1pt]{>}},
mark=at position 0.58 with {\arrow[line width=1pt]{>}},
mark=at position 0.75 with {\arrow[line width=1pt]{>}},
mark=at position 0.9 with {\arrow[line width=1pt]{>}}
}
]
\draw[help lines,->] (-4.5,0) -- (4,0) coordinate (xaxis);
\draw[help lines,->] (0,-2.2) -- (0,4.5) coordinate (yaxis);
\draw[fill] (0,1) circle (2pt);
\draw[fill] (0,-1) circle (2pt);
\draw[fill] (-1.2,.5) circle (2pt);
\draw[fill] (-1.2,-.5) circle (2pt);
\draw[color=blue,line width=0.8pt,->,decorate,decoration={zigzag,amplitude=2,segment length=9,post length=4}] (0,1) -- (0,3.8);
\draw[color=blue,line width=0.8pt,->,decorate,decoration={zigzag,amplitude=2,segment length=9,post length=4}] (0,-1) -- (0,-2);
\draw[color=blue,line width=0.8pt,->,decorate,decoration={zigzag,amplitude=2,segment length=9,post length=4}] (-1.2,.5) -- (-1.2,3.8);
\draw[color=blue,line width=0.8pt,->,decorate,decoration={zigzag,amplitude=2,segment length=9,post length=4}] (-1.2,-.5) -- (-1.2,-2);
\draw[line width=0.8pt,postaction=decorate] (0,0)  -- (3.5,0) arc (0:87:3.5) -- (0.2,0.95) arc (0:-180:0.2)  -- (-0.2,3.5) arc (93:106:3.5) -- (-1,.45) arc (0:-180:.2) -- (-1.4,3.3) arc (109:180:3.5) -- (0,0);
\draw[help lines] (3.5,3.5) -- (3.5,3.9);
\draw[help lines] (3.5,3.5) -- (4.1,3.5);
\node at (0.7,1) {$i\, |\vec{p}_1\,|$};
\node at (0.7,-1) {$-i\,|\vec{p}_1\,|$};
\node at (-2.5,0.5) {$-p^{\scriptscriptstyle{E}}_2 + \,  i \,  |\vec{p}_3|$};
\node at (-2.5,-0.6) {$-p^{\scriptscriptstyle{E}}_2 - \,  i \,  |\vec{p}_3|$};
\node at (3.8,3.7) {$p_1^{\scriptscriptstyle{E}}$};
\end{tikzpicture}
\caption{Closed contour of integration for the $p_1^{\scriptscriptstyle{E}}$ integral. The contribution from the arc at infinity vanishes, so the integral along the real axis is equal to that on each side of the two branch cuts on the upper-half plane.} \label{fig:contour-3pnt1}
\end{figure}
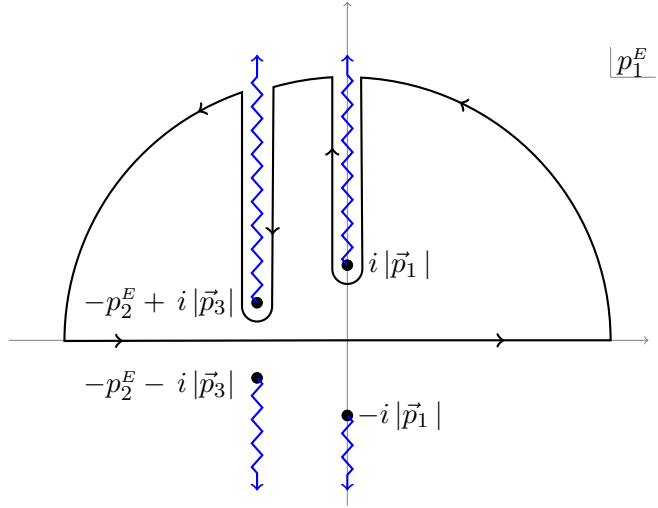

Since $\epsilon_{13} >0,$ we  consider a contour in the upper half going around the branch cuts, which extend from $i \, |\vec{p}_1|$  to $+i \, \infty$, and from $-p^{\scriptscriptstyle{E}}_2 + \,  i \,  |\vec{p}_1 + \vec{p}_2|$ to $-p^{\scriptscriptstyle{E}}_2 + \,  i \, \infty$ (see figure \ref{fig:contour-3pnt1}). 
The contributions from the arcs at infinity vanish because $\epsilon_{13} >0$. Since the integrand is analytic in the region inside the contour, the integral over the real axis equals that along each side of the branch cuts
\begin{align}
& \int\limits_{- \infty}^{\infty} \frac{d p^{\scriptscriptstyle{E}}_1}{2\pi} \, e^{-\,p^{\scriptscriptstyle{E}}_1 (t_{13} - i \, \epsilon_{13})} \, |p_1|^{\nu_1}\, |p_1+ p_2|^{\nu_3}\, K_{\nu_1}(|p_1| \, t) \,  K_{\nu_3}(|p_1+p_2|\, t) \notag \\[10pt]
 = \quad & \pi \int\limits_{|\vec{p}_1|}^{\infty} \frac{d p^{\scriptscriptstyle{0}}_1}{2\pi} \, e^{- i \,p^{\scriptscriptstyle{0}}_1 (t_{13} - i \, \epsilon_{13})} \, \left[(\,p^{\scriptscriptstyle{0}}_1)^2 - |\vec{p}_1|^2\right]^{\nu_1}\, \left[(i \,p^{\scriptscriptstyle{0}}_1 + p^{\scriptscriptstyle{E}}_2)^2 + |\vec{p}_1+ \vec{p}_2|^2\right]^{\nu_3/2} \notag \\[3pt]
 & \hspace{35mm} \times J_{\nu_1}\left(\sqrt{(\,p^{\scriptscriptstyle{0}}_1)^2 - |\vec{p}_1|^2}\, t\right) \,  K_{\nu_3}\left(\sqrt{(i \,p^{\scriptscriptstyle{0}}_1 + p^{\scriptscriptstyle{E}}_2)^2 + |\vec{p}_1+ \vec{p}_2|^2}\, t\right) \notag \\[7pt]
 \quad &  + \pi \int\limits_{|\vec{p}_1+ \vec{p}_2|}^{\infty} \frac{d p^{\scriptscriptstyle{0}}_1}{2\pi} \, e^{- ( i \, p^{\scriptscriptstyle{0}}_1 - p^{\scriptscriptstyle{E}}_2) (t_{13} - i \, \epsilon_{13})} \,  \left[(i \,p^{\scriptscriptstyle{0}}_1 - p^{\scriptscriptstyle{E}}_2)^2 + |\vec{p}_1|^2\right]^{\nu_1/2}\, \left[(p^{\scriptscriptstyle{0}}_1 )^2 - |\vec{p}_1+ \vec{p}_2|^2\right]^{\nu_3/2} \notag \\[3pt]
 & \hspace{35mm} \times K_{\nu_1}\left(\sqrt{(i \,p^{\scriptscriptstyle{0}}_1 - p^{\scriptscriptstyle{E}}_2)^2 + |\vec{p}_1|^2}\, t\right) \, J_{\nu_3}\left(\sqrt{(p^{\scriptscriptstyle{0}}_1 )^2 - |\vec{p}_1+ \vec{p}_2|^2} \, t\right),  \label{p1eint}
\end{align}
where in the first line $|p|$ is the Euclidean norm and we have used the identity\footnote{Note that we have chosen the branch cut for the $K$-Bessel function to run along the imaginary axis rather than the negative real line hence there's a subtlety in translating standard identities to our case. However, in most cases, such as here, the identities are unaffected by this issue.}
\begin{equation}
- \pi \, i \, J_{\nu}(z) = e^{\nu \pi i/2} K_{\nu}(z \, e^{i \pi/2}) - e^{- \nu \pi i/2} K_{\nu}(z\,  e^{- i \pi/2}) 
 \label{Jid}
\end{equation}
to rewrite the $K$-Bessel function in terms of the Bessel function of the first kind.

Note that, as in section \ref{sec:pre}, there are also contributions from the cups around the branch points (denoted $\C_c$ in figure \ref{fig:contour-2pnt}) but in this case these contributions vanish because 
\begin{equation}
 \lim_{\epsilon \rightarrow 0} \epsilon^{\nu+1} K_{\nu}(\epsilon) = 0.
\end{equation}

From Wick rotating $p^{\scriptscriptstyle{E}}_1$ we get two integrals, hence we need to consider two $p^{\scriptscriptstyle{E}}_2$ integrals:  
\begin{align}
&\int\limits_{- \infty}^{\infty} \frac{d p^{\scriptscriptstyle{E}}_2}{2\pi} \, e^{-\,p^{\scriptscriptstyle{E}}_2 (t_{23} - i \, \epsilon_{23})} \, |p_2|^{\nu_2} \, \left[(i \,p^{\scriptscriptstyle{0}}_1 + p^{\scriptscriptstyle{E}}_2)^2 + |\vec{p}_1+ \vec{p}_2|^2\right]^{\nu_3/2} \notag \\[3pt]
 & \hspace{45mm} \times K_{\nu_2}\left(|p_2|\, t \right) \,  K_{\nu_3}\left(\sqrt{(i \,p^{\scriptscriptstyle{0}}_1 + p^{\scriptscriptstyle{E}}_2)^2 + |\vec{p}_1+ \vec{p}_2|^2}\, t \right) \label{p2eint1}
 \end{align}
 with $p^{\scriptscriptstyle{0}}_1 \geq |\vec{p}_1|,$ coming from the first integral on the rhs of \eqref{p1eint}, 
 and
 \begin{align}
&\int\limits_{- \infty}^{\infty} \frac{d p^{\scriptscriptstyle{E}}_2}{2\pi} \, e^{p^{\scriptscriptstyle{E}}_2 (t_{12} - i \, \epsilon_{12})} \, \left[(i \,p^{\scriptscriptstyle{0}}_1 - p^{\scriptscriptstyle{E}}_2)^2 + |\vec{p}_1|^2\right]^{\nu_3/2}\, |p_2|^{\nu_2} \notag \\[3pt]
 & \hspace{45mm} \times K_{\nu_1}\left(\sqrt{(i \,p^{\scriptscriptstyle{0}}_1 - p^{\scriptscriptstyle{E}}_2)^2 + |\vec{p}_1|^2}\, t \right) \, K_{\nu_2}\left(|p_2|\, t\right)  \label{p2eint2}
\end{align}
 with $p^{\scriptscriptstyle{0}}_1 \geq |\vec{p}_1 + \vec{p}_2|,$ coming from the second integral on the rhs of \eqref{p1eint}. 
Note that the exponential in the second integral is slightly modified because of the $e^{ p^{\scriptscriptstyle{E}}_2 (t_{13} - i \, \epsilon_{13})}$ factor coming from the second integral in the Wick rotation of the $p^{\scriptscriptstyle{E}}_1$ integral, \eqref{p1eint}.

For the first integral, \eqref{p2eint1}, we again consider a closed contour in the upper half complex  $p^{\scriptscriptstyle{E}}_2$-plane going around the branch cuts. The integrand has branch points at 
\begin{equation}
 p^{\scriptscriptstyle{E}}_2 = \pm  i \, |\vec{p}_2|, \qquad p^{\scriptscriptstyle{E}}_2 = - i \, p^{\scriptscriptstyle{0}}_1 \pm i \, |\vec{p}_1 + \vec{p}_2|
\end{equation}
and at $ \infty$. We are only interested in branch points in the upper half plane, i.e.\ $ i \, |\vec{p}_2|$, and also possibly $- i \, p^{\scriptscriptstyle{0}}_1 + i \, |\vec{p}_1 + \vec{p}_2|$ depending on whether $|\vec{p}_1 + \vec{p}_2|> p^{\scriptscriptstyle{0}}_1$. In any case, using the triangle inequality and $p^{\scriptscriptstyle{0}}_1 \geq |\vec{p}_1|$, 
\begin{equation}
 - \, p^{\scriptscriptstyle{0}}_1 + \, |\vec{p}_1 + \vec{p}_2| \leq |\vec{p}_2|\,.
\end{equation}
Considering separately the two cases where there are one or two branch points on the upper half plane, the $p^{\scriptscriptstyle{E}}_2$ integral can be written in terms of an integral over $p^{\scriptscriptstyle{0}}_2$, as was done for $p^{\scriptscriptstyle{E}}_1$ in equation \eqref{p1eint}.

The second integral \eqref{p2eint2} can be treated in the same way. In this case though, because of the modification in the exponent and the fact that $\epsilon_{12}>0,$ we need to consider a contour in the lower half of the $ p^{\scriptscriptstyle{E}}_2$-plane. The branch points of the integrand in  \eqref{p2eint2} are at
\begin{equation}
 p^{\scriptscriptstyle{E}}_2 = \pm  i \, |\vec{p}_2|\,, \qquad p^{\scriptscriptstyle{E}}_2 = i \, p^{\scriptscriptstyle{0}}_1 \pm i \, |\vec{p}_1|
\end{equation}
and $\infty$. Since in this case $p^{\scriptscriptstyle{0}}_1 \geq |\vec{p}_1 + \vec{p}_2|$, using the triangle inequality
$$p^{\scriptscriptstyle{0}}_1 - |\vec{p}_1| > - |\vec{p}_2|\,.$$
Hence we again have two cases to consider depending on whether $p^{\scriptscriptstyle{0}}_1 - |\vec{p}_1|$ is positive or negative. 

Putting all these results together we rewrite the integral \eqref{FT-EL-3b}   as a Fourier transform, from which we identify the Lorentzian 3-point function in momentum space to be
\begin{align}
& C(p_1, p_2 ;\{\beta_j\})
= \frac{\pi^2}{2} \, \theta(p^{\scriptscriptstyle{0}}_1 - |\vec{p}_1|)\,  \theta(- p^{\scriptscriptstyle{0}}_3 - |\vec{p}_3|)   \, c_{\scriptscriptstyle{E}}(\{\beta_j\}) \,\int\limits_0^\infty dt \, t^{d/2-1}  \, \prod_{j=1}^{3}\, p_j^{\; \nu_j} \, \notag \\[7pt]
& \quad \hspace{22mm} \times \Bigg\{  2 \,\theta(p^{\scriptscriptstyle{0}}_2 + |\vec{p}_2|)  \, \theta(|\vec{p}_2|- p^{\scriptscriptstyle{0}}_2  ) \,  J_{\nu_1}(p_1 \, t) \,  K_{\nu_2}(p_2 \, t) \, J_{\nu_3}(p_3 \, t) \notag \\[6pt]
& \hspace{32mm}-\pi \, \theta(p^{\scriptscriptstyle{0}}_2 - |\vec{p}_2|) \,  J_{\nu_1}(p_1 \, t)\, \left[J_{\nu_2}(p_2 \, t) \, Y_{\nu_3}(p_3 \, t) + Y_{\nu_2}(p_2 \, t) \, J_{\nu_3}(p_3 \, t) \right] \notag\\
& \hspace{32mm}-\pi \, \theta(-p^{\scriptscriptstyle{0}}_2 - |\vec{p}_2|) \,   \left[J_{\nu_1}(p_1 \, t) \, Y_{\nu_2}(p_2 \, t) + Y_{\nu_1}(p_1 \, t) \, J_{\nu_2}(p_2 \, t) \right] \,  J_{\nu_3}(p_3 \, t)\Bigg\}, \label{3BL}
\end{align}
where on the rhs $p_j$ denotes the Lorentzian norm $\sqrt{|(p^{\scriptscriptstyle{0}}_j)^2 - |\vec{p}_j|^2 | }$; $p_3 = -( p_1+p_2)$ as a $d$-vector; the coefficient $c_{\scriptscriptstyle{E}}$ is defined in \eqref{cEdef}; and $Y_{\nu}(z)$ is the Bessel function of the second kind. 
In deriving the above expression we have used 
\begin{equation}
 e^{-\nu \pi i/2} K_{\nu}(e^{-i \pi/2} z) = \frac{ i \, \pi}{2} \left(J_{\nu}(z)+ i \, Y_{\nu}(z)\right)\,, \quad z \in \mathbb{R}\,, 
\end{equation}
and its complex conjugate.  

In comparison to the Euclidean 3-point function \eqref{3KE}, it is clear that the Lorentzian one is more complicated because it has to incorporate complicated causal relations which are partly achieved through the Heaviside step functions. 

A consistency check can be found by complex conjugating the definition of $C(p_1, p_2 ;\{\beta_j\})$, 
\begin{align} \label{Cdefcc}
\langle \mathcal{O}_1(x_1)\,\mathcal{O}_2(x_2)\,\mathcal{O}_3(x_3)\rangle^{*}
=\int \frac{d^dp_1}{(2\pi)^d} \frac{d^dp_2}{(2\pi)^d}\, e^{-i\,p_1\cdot x_{13}}\,e^{-i\,p_2\cdot x_{23}}\,C(p_1, p_2 ;\{\beta_j\})^{*}\,.
\end{align}
First let us consider the Euclidean case. The Euclidean correlator in position space is real, hence reparametrising the momenta integrals by letting $p_j \rightarrow - p_j$, we find that 
\begin{equation}
 C_{\scriptscriptstyle{E}}(p_1, p_2 ;\{\beta_j\})^{*} = C_{\scriptscriptstyle{E}}(-p_1, -p_2 ;\{\beta_j\})\,.
\end{equation}
We can verify that the expression given in equation \eqref{3KE} indeed satisfies the consistency condition above, for $C_{\scriptscriptstyle{E}}$ is real and only depends of the norm of the momenta. 

In the Lorentzian case, the consistency condition is complicated by the fact that the correlators in position space are not strictly real because of the $i \, \epsilon$ prescription that specifies the order of operators in the correlation function. In fact 
\begin{equation}
 \langle \mathcal{O}_1(x_1)\,\mathcal{O}_2(x_2)\,\mathcal{O}_3(x_3)\rangle^{*} =  \langle \mathcal{O}_3(x_3)\,\mathcal{O}_2(x_2)\,\mathcal{O}_1(x_1)\rangle\,.
\end{equation}
Hence equation \eqref{Cdefcc} becomes
\begin{align} 
\langle \mathcal{O}_3(x_3)\,\mathcal{O}_2(x_2)\,\mathcal{O}_1(x_1)\rangle
=\int \frac{d^dp_1}{(2\pi)^d} \frac{d^dp_2}{(2\pi)^d}\, e^{i\,(p_1 + p_2)\cdot x_{31}}\,e^{- i\,p_2\cdot x_{21}}\,C(p_1 , p_2 ;\{\beta_j\})^{*}\,,
\end{align}
where on the rhs we have rearranged the terms so that the $x_j$-dependence of the exponentials are such that they define an appropriate Fourier transform of the correlator on the lhs. Therefore, from the above equation the consistency condition on the Lorentzian 3-point function in momentum space is 
\begin{equation}
 C(p_1, p_2 ;\beta_1, \beta_2, \beta_3)^{*} = C(p_1+ p_2, - p_2 ; \beta_3, \beta_2, \beta_1)\,.
\end{equation}
The Lorentzian 3-point function \eqref{3BL} is real and it is straightforward to verify that it satisfies the above relation. 

\subsubsection{Finiteness of the Lorentzian 3-point function}
\label{sec:div}

We now analyse possible divergences of this triple-Bessel expression, and start with those coming from the upper limit of the $t$-integral. 
Given that the modified Bessel function $K_\nu(z)$ behaves as a decaying exponential at large $z$ (see \eqref{Bessels-large-z}), the first integral $J_{\nu_1}K_{\nu_2}J_{\nu_3}$ does not have a divergence coming form the large $t$ region. Instead, the $JYJ$ integrals do because 
both $J_\nu(z)$ and $Y_\nu(z)$ only decay as a power law. It follows in fact that the $JYJ$ integrals only exhibit large-$t$ convergence if $d<5$, since the large $t$ contribution is of the form
\begin{equation}
\int\limits^\infty dt\, t^{\frac{d}{2}-\frac{5}{2}} \,e^{i t}.
\end{equation}
However, the expression above is given by  ${}_1 F_{2}( { \scriptstyle{- \frac{1}{4}}} )$ generalised hypergeometric functions which are entire. Hence the integral can be analytically continued to all $d$.

Just as in the case of the Euclidean triple-K expression, the $t$-integral in \eqref{3BL} can diverge for certain values of the $\beta_j$ due to the behaviour of the integrand at small $t$. Indeed, taking into account the expansions  of the different Bessel functions, \eqref{app:conv-Bessels}, involved,
the triple-Bessel expression only converges if
\begin{equation}\label{conv-condL}
\frac{d}{2}>\text{sup}\{ |\nu_1|-\nu_2-\nu_3,|\nu_2|-\nu_1-\nu_3,|\nu_3|-\nu_1-\nu_2\}.
\end{equation}
Note that this is weaker than the analogous condition in the Euclidean case \eqref{conv-condE}. If the above is not satisfied, the integral in principle diverges, but it can be defined by analytical continuation, as explained in \cite{BMS:imp}. However, from the expansion of the Bessel functions  \eqref{app:conv-Bessels} follows that the analytic continuation of  the $J_{\nu_1}K_{\nu_2}J_{\nu_3}$ integral fails if 
\begin{equation}
d/2+\nu_1\pm\nu_2+\nu_3=-2n, \qquad n\in\mathbb{N}^0.
\end{equation}
A similar reasoning follows for the  $JYJ$ (and permutations of $Y$) integrals in \eqref{3BL}. So the $t$-integral is divergent if either of the following conditions hold
\begin{equation}\label{div-condL}
\frac{d}{2}+\nu_t=-2n\,\qquad \text{or} \qquad \frac{d}{2}+\nu_t-2\nu_j=-2n, \qquad n\in\mathbb{N}^0.
\end{equation}
If we compare these conditions with those for the triple-K expression \eqref{div-condition}, 
\begin{equation}
d/2\pm \nu_1\pm\nu_2\pm\nu_3 =-2n,
\end{equation}
 we can see that the former are a subset of the latter, where only the sign choices with $(+++)$ and $(++-)$ and permutations are included. The reason is that in the Lorentzian case, each term of the triple-Bessel expression contains at least two $J_\nu$ functions. Since $J_\nu (z) \sim z^\nu$ for small $z$, this ensures that the divergence condition contains at least two $+$ signs. 

This difference is important, because \eqref{div-condL} is equivalent to
\begin{equation}
\Delta_t-d=-2 n \qquad \text{or} \qquad \beta_j=-2n ;
\end{equation}
in either case the coefficient $c_{\scriptscriptstyle{E}}(\{\beta_j\})$ vanishes due to the $\Gamma$-functions in the denominator. This implies that in all cases when the $t$-integral diverges, the coefficient vanishes, and therefore a regularisation can be used to render the triple-Bessel expression finite, hence requiring no renormalisation (see next section for an example of this). This is an important distinction with the Euclidean case, where as argued below \eqref{conv-condE}, in the cases with the $(---)$ and $(+--)$ sign options, the integral diverges and the coefficient is finite, hence the 3-point function requires renormalisation \cite{BMS:renoms}.

\subsection{Check}
\label{sec:check}

In the Euclidean case, it can be shown that the momentum-integrated expression of the Euclidean 3-point function \eqref{E3pt} is equivalent to the triple-$K$ form \eqref{3KE} \cite{BMS:imp}. We have Wick rotated each expression to obtain the Lorentzian 3-point function as a momentum integral,  section \ref{sec:scam}, and as a triple-Bessel expression, expression \eqref{3BL}. As a check, however,  we consider the 3-point function for scalars of dimension $\Delta_{j} = 1$, which implies $\beta_{j}= - \nu_{j} = 1/2$,  in $d=3$.

The Euclidean 3-point function in this case, from either equation \eqref{E3pt} or \eqref{3KE}, is \cite{BMS:imp}
\begin{equation}
 C_{\scriptscriptstyle{E}}(\{1/2\}) = \frac{1}{p_1 p_2 p_3},
\end{equation}
where we have chosen $c_{123} =(2 \pi )^{-3}.$ Directly Wick rotating this expression to Lorentzian space using the method outlined in this paper, the Lorentzian correlator is 
\begin{equation} \label{eg3pt}
 C(\{1/2\}) = \frac{4}{|p_1| \, |p_2| \, |p_3| } \theta(p^{\scriptscriptstyle{0}}_1 - |\vec{p}_1|)\,  \theta(- p^{\scriptscriptstyle{0}}_3 - |\vec{p}_3|)   \,\theta(p^{\scriptscriptstyle{0}}_2 + |\vec{p}_2|)  \, \theta(|\vec{p}_2|- p^{\scriptscriptstyle{0}}_2  ).
\end{equation}

We now consider the momentum-integrated expression. The case we are considering of $\beta_j = 1/2$ corresponds to a special case where the three 2-point functions being integrated are of the case II type. Using equations \eqref{3pt} and \eqref{k-integral-sp}, the momentum-integrated 3-point function is given by
\begin{equation}
C(\{1/2\})= \int d^3 k\,
\frac{\delta(p_{2}^{\scriptscriptstyle{0}}+k^{\scriptscriptstyle{0}}-|\vec{p}_2+\vec{k}|)\,\delta(p_1^{\scriptscriptstyle{0}}-k^{\scriptscriptstyle{0}}-|\vec{p}_1-\vec{k}|)\,\delta(k^{\scriptscriptstyle{0}}-|\vec{k}|)}{|\vec{p}_2+\vec{k}|\,|\vec{p}_1-\vec{k}|\,|\vec{k}|}.
\end{equation}
A straightforward, yet tedious, calculation confirms that the above integral agrees with \eqref{eg3pt}. For details on this computation see appendix \ref{sec:app-check}.

Finally, we consider the triple-Bessel expression \eqref{3BL} for $\nu_{j}=-1/2$, which is not well-defined since it satisfies the divergent condition \eqref{div-condL} and requires regularisation. This can be done by letting
\begin{equation}
 d \rightarrow d + 2 \epsilon, \qquad \Delta_{j} \rightarrow \Delta_{j} + \epsilon.
\end{equation}
This regularisation is convenient because it regularises the power of $t$ in the integrand but leaves the indices $\nu_j$ of the Bessel functions intact.

Using \begin{align}
 J_{-1/2}(x) = \sqrt{\frac{2}{\pi x}} \cos{x}, \qquad  Y_{-1/2}(x) = \sqrt{\frac{2}{\pi x}} \sin{x}, \qquad  K_{-1/2}(x) = \sqrt{\frac{\pi}{2 x}} e^{-x},
\end{align}
and the regularisation above we find that the $J_{\nu_1} K_{\nu_2} J_{\nu_3}$ integral of \eqref{3BL} acquires a  $1/\epsilon$ pole, while the other $JJY$ integrals are actually finite. When multiplied with the $c_{\scriptscriptstyle{E}}(\{\beta_j\})$ coefficient \eqref{cEdef}, which is proportional to
$\epsilon$, only the former contributes, and \eqref{3BL} reduces to \eqref{eg3pt}.

\section{Tensorial correlators}
\label{sec:ten}

There are at least two ways to calculate tensorial correlators in momentum space. The first is to write down a tensorial decomposition in terms of the momenta that satisfy the properties of the tensors in the correlator---for example conservation or tracelessness. The form factors are then found by solving conformal Ward identities. This is the method that is used in Ref.~\cite{BMS:imp} in order to find Euclidean 3-point correlators. The form factors in the Euclidean case are given by triple $K$-integrals of the form \eqref{3KE} with general exponents for the variables in the integral that are determined in each case. The advantage of this approach is that the properties of the tensors in the correlator are explicit and all correlators are given by the same basic building blocks, namely the triple $K$-integral.    

An alternative approach to find tensorial correlators is to directly use the 3-point function of scalars. As we explain below, since tensorial correlators in position space are given by tensorial structures involving positions, tensorial correlators in momentum space are then just momentum derivatives of scalar correlators. The advantage of this approach is that once the scalar correlator is known, it is straightforward, in principle, to find all the tensorial correlators. The disadvantage is that the properties of tensors in the correlation functions, such as tracelessness, are obscured and are not explicit. For example, in the Euclidean case this approach leads to correlators that are not immediately the same as the tensorial correlators in Ref.~\cite{BMS:imp}. They are related to each other up to boundary terms, which is subtle when the triple-$K$ integral is divergent. 

Of course, we can also Wick rotate the tensorial correlators in Ref.~\cite{BMS:imp}. However, already for the scalar case, the Lorentzian triple Bessel function form \eqref{3BL} is significantly more involved than the Euclidean triple-$K$ integral. Hence we will take the second approach outlined above to find tensorial 3-point correlators using the scalar correlator.

By translation invariance the tensorial part of correlators with tensors is given by the differences of the position vectors $x_1^{\mu}, x_2^{\mu}, x_3^{\mu}.$ Namely a general correlator will be a sum of terms of the form 
\begin{equation} \label{genten}
\frac{x_{ij}^{\mu_1} \,\,x_{kl}^{\mu_2}\, \dots }{(x_{23}^2)^{\beta_1}\,(x_{13}^2)^{\beta_2}\,(x_{12}^2)^{\beta_3}}\,,
\end{equation}
where $i \neq j, k \neq l \in \{1,2,3\},$ as can be verified by referring to the expressions in  Ref.~\cite{OP}. 
Consider again the Fourier transform of the Lorentzian 3-point function of scalar operators, \eqref{mcorreldef},\footnote{In all of this section, the $C(p_1, p_2; \{\beta_j\})$ function and the coefficients $c(\{\beta_j\})$ do not include the 3-point function coefficient $c_{123}$.}
\begin{equation} \label{Cdef}
\int \prod_{j=1}^3 \left( d^dx_j \,  e^{- i p_j \cdot x_j} \right) \frac{1}{(x_{23}^2)^{\beta_1}\,(x_{13}^2)^{\beta_2}\,(x_{12}^2)^{\beta_3}}= (2 \pi)^d \, \delta^{(d)}(p_1 + p_2 +p_3) \,  C(p_1, p_2; \{\beta_j\}) \,.
\end{equation}
When we write the Fourier transform of a tensorial correlator, the $x_{ij}^{\mu_1}$ from expression \eqref{genten} becomes a difference of partial derivatives with respect to the momenta $p_i$ and $p_j$, which then act on the scalar Fourier transform above. On the rhs, the momentum preserving $\delta$-function commutes with the difference of partial derivatives, hence we need just consider derivatives of the $\delta$-function stripped correlator $C(p_1, p_2;\{\beta_j\})$.

For example consider the following integral
\begin{align} \label{teneg}
\int \prod_i d^d & x_i \,e^{- i p_i\cdot x_i}  \frac{x_{12 \; \mu}  x_{12 \; \nu} }{(x_{23}^2)^{\beta_1}\,(x_{13}^2)^{\beta_2}\,(x_{12}^2)^{\beta_3}} \notag \\[7pt]
 &=  - (2 \pi)^d \delta(p_1 + p_2 + p_3) \left( \frac{\del}{\del p_1^{\mu}} - \frac{\del}{\del p_2^{\mu}} \right) \left( \frac{\del}{\del p_1^{\nu}} - \frac{\del}{\del p_2^{\nu}} \right) C(p_1, p_2; \{\beta_j\})
\end{align}
 with $ \beta_3 > d/2+1$.

Consider first 
\begin{equation}
 \left( \frac{\del}{\del p_1^{\nu}} - \frac{\del}{\del p_2^{\nu}} \right) C(p_1, p_2; \{\beta_j\}).
\end{equation}
Using expression  \eqref{eq1} for $C(p_1, p_2; \{\beta_j\})$, and reparametrising the $k$-integral as $k \rightarrow k-p_2$, only one derivative needs to be computed,  and this becomes equal to 
\begin{equation}\label{first-derivative}
(2\, \beta_3-d)  c(\{\beta_j\}) \int \frac{d^dk}{(2\pi)^d} 
\frac{\theta(k^{\scriptscriptstyle{0}}-|\vec{k}|)\,
\theta(p_1^{\scriptscriptstyle{0}}+p_2^{\scriptscriptstyle{0}}-k^{\scriptscriptstyle{0}}-|\vec{p}_1+\vec{p}_2-\vec{k}|)\,\theta(k^{\scriptscriptstyle{0}}-p_2^{\scriptscriptstyle{0}}-|\vec{k}-\vec{p}_2|)}{|k|^{d - 2 \beta_1}\,|p_1+p_2-k|^{d - 2 \beta_2}\,|k-p_2|^{d - 2 \beta_3+2}} (p_{2}-k)_{\nu} \, ,
\end{equation}
where notice that
\begin{equation}
|k-p_2|^2 = - \, ( k-p_2)^2
\end{equation}
and the derivative of the Heaviside step function 
\begin{equation}\label{derivative-theta}
 \partial_{p_2^{\mu}} \theta(k^{\scriptscriptstyle{0}}-p_2^{\scriptscriptstyle{0}}-|\vec{k}-\vec{p}_2|) =   (k-p_2)_{\mu} \, \frac{\delta(k^{\scriptscriptstyle{0}}-p_2^{\scriptscriptstyle{0}}-|\vec{k}-\vec{p}_2|)}{|\vec{k}-\vec{p}_2|}\,,
\end{equation}
gives a vanishing contribution because of the factor of $(k-p_2)^2$ with a positive power. If we had not chosen  $\beta_3$  large enough, this derivative would also contribute.

Acting next with $-(\partial_{p_1^{\mu}} -\partial_{p_2^{\mu}})$ on \eqref{first-derivative}, and reparametrising $k\rightarrow k+p_2$ gives
\begin{align}\label{egans}
- (2\, \beta_3-d) \,  c(\{\beta_j\})& \int \frac{d^dk}{(2\pi)^d} \,
\frac{\theta(p_2^{\scriptscriptstyle{0}}+k^{\scriptscriptstyle{0}}-|\vec{p}_2+\vec{k}|)\,
\theta(p_1^{\scriptscriptstyle{0}}-k^{\scriptscriptstyle{0}}-|\vec{p}_1-\vec{k}|)\,\theta(k^{\scriptscriptstyle{0}}-|\vec{k}|)}{|p_2+k|^{d - 2 \beta_1}\,|p_1-k|^{d - 2 \beta_2}\,|k|^{d - 2 \beta_3}} \notag \\[11pt]
&\hspace{50mm}\times \frac{(2\beta_3-2-d) k_{\mu}k_{\nu}-|k|^2\,\eta_{\mu\nu}}{| k |^4} \, .
\end{align}
Hence this is the $\delta$-function stripped Fourier transform on the lhs of equation \eqref{teneg}.

As a consistency check, we take the trace of the above expression \eqref{egans}, and obtain
\begin{align}
c(\beta_1,\beta_2,\beta_3-1) \int \frac{d^dk}{(2\pi)^d} \,
\frac{\theta(p_2^{\scriptscriptstyle{0}}+k^{\scriptscriptstyle{0}}-|\vec{p}_2+\vec{k}|)\,
\theta(p_1^{\scriptscriptstyle{0}}-k^{\scriptscriptstyle{0}}-|\vec{p}_1-\vec{k}|)\,\theta(k^{\scriptscriptstyle{0}}-|\vec{k}|)}{|p_2+k|^{d - 2 \beta_1}\,|p_1-k|^{d - 2 \beta_2}\,|k|^{d - 2 (\beta_3-1)}} \, ,
\end{align}
confirming that
\begin{equation}\label{consistency}
\eta^{\mu\nu}\int \prod_i d^d x_i \,e^{- i p_i\cdot x_i}  \frac{x_{12 \; \mu}  x_{12 \; \nu} }{(x_{23}^2)^{\beta_1}\,(x_{13}^2)^{\beta_2}\,(x_{12}^2)^{\beta_3}}
= \int \prod_i d^d  x_i \,e^{- i p_i\cdot x_i}  \frac{1 }{(x_{23}^2)^{\beta_1}\,(x_{13}^2)^{\beta_2}\,(x_{12}^2)^{\beta_3-1}} \, .
\end{equation}

\subsection{$\langle \mathcal{O} T_{\mu \nu} \mathcal{O} \rangle$ correlator} 

Having explained the general method for calculating tensorial correlators in the previous section, we apply the procedure to a correlator of two identical scalar operators $\mathcal{O}$ and an energy-momentum tensor $T_{\mu \nu}$. This correlator is used in the next section to calculate the expectation value of the ANEC operator in a state created by the operator $\mathcal{O}.$ 

In position space \cite{OP}, 
\begin{equation}
 \langle \mathcal{O}(x_1) T_{\mu \nu}(x_2) \mathcal{O}(x_3) \rangle = \frac{a}{x_{23}^{d-2}\,x_{13}^{2 \, \Delta - d+2}\,x_{12}^{d-2}} \left(  \left( \frac{x_{12\; \mu}}{x_{12}^2} - \frac{x_{32\; \mu}}{x_{32}^2} \right)\left( \frac{x_{12\; \nu}}{x_{12}^2} - \frac{x_{32\; \nu}}{x_{32}^2} \right) - \frac{1}{d} \frac{x_{13}^2}{x_{12}^2 \,x_{23}^2} \eta_{\mu \nu} \right) 
\end{equation}
where 
\begin{equation}
 x_{ij}^{\mu} =  (t_{ij} - i \epsilon_{ij}, \vec{x}_{ij} )
\end{equation}
and $\epsilon_{ij} > 0$ for $i<j$, otherwise negative.  

The $\delta$-function stripped correlator in momentum space is then
\begin{align}
 \llangle \mathcal{O}(p_1) T_{\mu \nu}(p_2) \mathcal{O}(p_3) \rrangle &= a \left( \frac{\del}{\del p_2^{\mu}} - \frac{\del}{\del p_1^{\mu}} \right) \left( \frac{\del}{\del p_1^{\nu}} - \frac{\del}{\del p_2^{\nu}} \right) C(p_1, p_2; \frac{d-2}{2}, \Delta+1 -\frac{d}{2}, \frac{d+2}{2}) \notag \\[5pt]
& \quad + 2\, a \left( \frac{\del}{\del p_2^{(\mu}} - \frac{\del}{\del p_1^{(\mu}} \right) \frac{\del}{\del p_2^{\nu)}} \,  C(p_1, p_2; \frac{d}{2},\Delta+1 -\frac{d}{2}, \frac{d}{2}) \notag \\[5pt]
& \quad - a \,  \frac{\del}{\del p_2^{\mu}}  \, \frac{\del}{\del p_2^{\nu}} \,  C(p_1, p_2;\frac{d+2}{2}, \Delta+1 -\frac{d}{2}, \frac{d-2}{2}) \notag \\[5pt]
& \quad - \frac{a}{d} \, \eta_{\mu \nu} \,  C(p_1, p_2; \frac{d}{2}, \Delta-\frac{ d}{2}, \frac{d}{2}), \label{iexpoto}
\end{align}
where we have used the fact that the $C$ have no dependence on $p_3$. Using the results of section \ref{sec:scam}, 
\begin{align}
 C(\frac{d-2}{2}, \Delta+1 -\frac{d}{2}, \frac{d+2}{2}) &=
\frac{c}{a}
 \int \frac{d^dk}{(2\pi)^d}\,
\frac{\theta(p_1^{\scriptscriptstyle{0}}-k^{\scriptscriptstyle{0}}-|\vec{p}_1-\vec{k}|)\,\theta(k^{\scriptscriptstyle{0}}-|\vec{k}|)}{|p_1-k|^{2 (d-1 - \Delta)}\,|k|^{- 2}}\frac{\delta(p_2^{\scriptscriptstyle{0}}+k^{\scriptscriptstyle{0}}-|\vec{p}_2+\vec{k}|)}{|\vec{p}_2+\vec{k}|}, \label{t1}
\end{align}
\begin{align}
 C(\frac{d+2}{2}, \Delta+1 -\frac{d}{2},\frac{d-2}{2}& )=\frac{c}{a} \int \frac{d^dk}{(2\pi)^d}\,
\frac{\theta(p_1^{\scriptscriptstyle{0}}-k^{\scriptscriptstyle{0}}-|\vec{p}_1-\vec{k}|)\,\theta(p_{2}^{\scriptscriptstyle{0}} + k^{\scriptscriptstyle{0}}-|\vec{p_2}+\vec{k}|)}{|p_1-k|^{2 (d-1 - \Delta)}\,|p_2+k|^{- 2}} \frac{\delta(k^{\scriptscriptstyle{0}}-|\vec{k}|)}{|\vec{k}|},\label{t3}
\end{align}
\begin{align}
 C(\frac{d}{2}, \Delta+1 -\frac{d}{2}, \frac{d}{2})&=
\frac{ 2 \, d \, c}{(d-2)\, a} \int \frac{d^dk}{(2\pi)^d}\,
\frac{\theta(p_{2}^{\scriptscriptstyle{0}}+k^{\scriptscriptstyle{0}}-|\vec{p}_2+\vec{k}|)\,\theta(p_1^{\scriptscriptstyle{0}}-k^{\scriptscriptstyle{0}}-|\vec{p}_1-\vec{k}|)\,\theta(k^{\scriptscriptstyle{0}}-|\vec{k}|)}{\,|p_1-k|^{2(d -1- \Delta)}},\label{t2}
\end{align}
where
\begin{equation}
 c= \frac{ 4^{d-1-\Delta} \, \pi^{3(d/2+1)} \, d(d-2)\, a}{\Gamma(\Delta+1 -d/2)\,\Gamma(\Delta+2 - d)\,\Gamma(d/2+1)^2}.
 \label{cotodef}
\end{equation}

We note that $C(\frac{d-2}{2}, \Delta+1 -\frac{d}{2}, \frac{d+2}{2})$ has a $\delta(p_{2}^{\scriptscriptstyle{0}}+k^{\scriptscriptstyle{0}}-|\vec{p}_2+\vec{k}|)$. Reparametrising the $k$-integral by letting $k \rightarrow k-p_2$ as we did in the general tensorial example, turns the argument of the $\delta$-function into $k^{\scriptscriptstyle{0}}-|\vec{k}|$, hence becoming independent of $p_1$ and $p_2$. This allows us to avoid derivatives of the Dirac $\delta$-function when evaluating the rhs of \eqref{iexpoto}. 

Using identity \eqref{derivative-theta},
it is straightforward to show that
\begin{align}
  \llangle \mathcal{O}(p_1) T_{\mu \nu}(p_2) \mathcal{O}(p_3) \rrangle &=   -4 \, c  \int \frac{d^dk}{(2\pi)^d}\,
\frac{\theta(k^{\scriptscriptstyle{0}})\,\theta(p_{2}^{\scriptscriptstyle{0}}+k^{\scriptscriptstyle{0}})\,\theta(p_1^{\scriptscriptstyle{0}}-k^{\scriptscriptstyle{0}}-|\vec{p}_1-\vec{k}|)}{\,|p_1-k|^{2(d -1- \Delta)}} \notag \\[9pt]
& \hspace{10mm} \times \Bigg( \left[ \frac{8(d-1)}{d-2} \,  \left(k_{\mu} k_{\nu} +  \, k_{(\mu} p_{2 \; \nu)} \right) + 2 \, p_{2 \; \mu} p_{2 \; \nu} \right]\,  \delta(k^2)  \,\delta((p_2 + k)^2)\,  \notag \\[7pt]
& \hspace{18mm} -  \, \theta(k^{\scriptscriptstyle{0}}-|\vec{k}|) \,\delta((p_2 + k)^2)\, \eta_{\mu \nu}  -  \, \theta(p_2^{\scriptscriptstyle{0}}+k^{\scriptscriptstyle{0}}-|\vec{p}_2+\vec{k}|) \,\delta(k^2)\, \eta_{\mu \nu}  \notag \\[16pt]
& \hspace{18mm} +  \frac{(2 \Delta-d) (\Delta-d+1)}{d-2} \, \frac{\theta(k^{\scriptscriptstyle{0}}-|\vec{k}|) \,  \theta(p_2^{\scriptscriptstyle{0}}+k^{\scriptscriptstyle{0}}-|\vec{p}_2+\vec{k}|)}{|p_1-k|^2} \, \eta_{\mu \nu} \Bigg),\label{OTO-d}
\end{align}
where we have used 
\begin{equation}
 \theta(p^{\scriptscriptstyle{0}})\delta(p^2)=\frac{\delta(p^{\scriptscriptstyle{0}}-|\vec{p}\,|)}{2 \,|\vec{p}\,|}.
\end{equation}

Equation \eqref{OTO-d} is valid for any dimension except for $d=2$ dimensions. This is because in that case, the first and third $C$-functions  required in \eqref{iexpoto} are not of the case II type, but rather of the case I type of section \ref{sec:scam}, and hence are not given by  \eqref{t1}, \eqref{t3}. The $d=2$ correlator can then be computed by using \eqref{L-int} instead. 

Note that the expression we have given, \eqref{OTO-d}, is not manifestly traceless or transverse. In fact,
\begin{equation}
 p_{2}^{\nu} \, \llangle \mathcal{O}(p_1) T_{\mu \nu}(p_2) \mathcal{O}(p_3) \rrangle = \frac{p_{2 \, \mu}}{d} \llangle \mathcal{O}(p_1) T^{\mu}{}_{\mu}(p_2) \mathcal{O}(p_3) \rrangle.
\end{equation}
As we have already mentioned, when the tensorial properties are not manifest, integration by parts is required to show them. When $\Delta > d-1,$ we can show that the trace is proportional to a total derivative, \emph{viz}.
\begin{align}
\int \frac{d^dk}{(2\pi)^d}\, & \eta^{\mu \nu} \frac{\partial}{\partial k^{\mu}} \left(\frac{\partial}{\partial k^{\nu}} + 2\, \frac{\partial}{\partial p^{\nu}}\right)
\frac{\theta(k^{\scriptscriptstyle{0}})\,\theta(p_{2}^{\scriptscriptstyle{0}}+k^{\scriptscriptstyle{0}})\,\theta(p_1^{\scriptscriptstyle{0}}-k^{\scriptscriptstyle{0}}-|\vec{p}_1-\vec{k}|)}{\,|p_1-k|^{2(d -1- \Delta)}}.
\end{align}
Hence the correlator is traceless and transverse. When $\Delta \leq d-1$ the analysis becomes more involved because the derivative of the step function $\theta(p_1^{\scriptscriptstyle{0}}-k^{\scriptscriptstyle{0}}-|\vec{p}_1-\vec{k}|)$ leads to terms the form $x^a \delta(x)$ for $a<0$ which must be regularised. 

In the example at the start of this section, we calculated the rhs of expression \eqref{teneg} by first reparametrising the integration variable $k$; had we not done this we would have obtained a more complicated expression for which the trace consistency check would not have been straightforward. In fact, to prove expression \eqref{consistency} in that case would have required integration by parts. This illustrates the point made at the start of this section that tensorial identities are not guaranteed to be manifest in this method and that one may have to use integration by parts in order to prove them. 

The $\mathcal{O}T\mathcal{O}$ correlator computed satisfies the transverse and trace Ward identities without contact terms, consistent with the $i \epsilon$ prescription in position space. We leave the implications of this for the interpretation of anomalies in Lorentzian signature for future work.

We can use the same methods outlined in this section to calculate other tensorial correlators. While the procedure in all cases is simple, technically the calculation does become quite complicated if there are a lot of tensors. It is foreseeable that there is a structure underlying these expressions that lends itself to generalisation; such a structure would, for example, make the tensorial properties of the correlators manifest. It may also be possible that the triple Bessel form of the 3-point function can be used to write down tensorial correlators systematically, as in \cite{BMS:imp}. We wish to report on this direction in a separate manuscript.

\section{ANEC expectation values on Hofman-Maldacena states}
\label{sec:Hofmal}

As an application of the expression obtained in the previous section for $\langle\mathcal{O}\,T_{\mu\nu}\,\mathcal{O}\rangle$ we use it here to reproduce the positivity of the  ANEC operator when evaluated on the simplest type of the Hofman-Maldacena states, namely those generated by scalar operators~\cite{HofMal, Hofman:2016awc, Hartman:2016lgu}, but in momentum space. Also, we generalise this to general dimension $d$.

 In the conformal-collider experiments \cite{HofMal}, one places calorimeters on a 2-sphere at infinity. These calorimeters measure the energy produced by some CFT operator $\mathcal{O}$ localised around the center of the sphere. The measurement of a calorimeter is then given by the expectation value of the time-integrated energy-momentum tensor on a state created by the operator $\mathcal{O}$, hence by  
a conformal 3-point function of the type $\langle \mathcal{O} \int T\, \mathcal{O}\rangle$. In fact, since the energy-momentum tensor is inserted at infinity, the integral over time of its energy component becomes the ANEC operator,
\begin{equation}
\mathcal{E}=\int dx^- \,T_{--}.
\end{equation}
 
Given that conformal 3-point correlators are completely fixed up to constants, requiring positivity of the integrated energy measurement places bounds on these constants. Concretely, when the normalised expectation value is computed on a state created by a scalar operator, the constant is simply a numerical coefficient, and is indeed positive. When the normalised expectation value is computed on a state created by the energy-momentum tensor itself, the constants depend on the conformal anomalies, and thus in four dimensions the positivity condition puts bounds on the $a$ and $c$ central charges.

To compute the bounds, Hofman-Maldacena used a particular type of states, namely wave packets with purely-timelike momentum
\begin{equation}\label{HM-state}
| \mathcal{O}(q)\rangle\equiv \int d^dx \, e^{-i q t}\,e^{-\frac{t^2+\vec{x}^2}{\sigma^2}}\, \mathcal{O}(x)\,|0\rangle\, .
\end{equation}
They consider them in the regime $q\,\sigma>>1$, which ensures that the operator has finite norm and is localised near the origin, which can be understood as a requirement for IR finiteness. In the calculation however, the Gaussian factor can be dropped. As pointed out in Ref.~\cite{Hartman:2016lgu}, this can lead to unphysical divergences, but these can be regularised with dimensional regularisation, we disregard this issue in the following.

The Hofman-Maldacena states yield optimal bounds for the ratios of the two anomaly coefficients \cite{Hartman:2016lgu}, because of the fact that the ANEC operator is inserted at null infinity. There, it commutes with the momentum operator $[\mathcal{E},P]=0$, hence momentum eigenstates are also eigenstates of the ANEC operator.\footnote{We thank Alexander Zhiboedov for pointing this out to us.} As a consequence, $\langle\mathcal{E}\rangle$ becomes diagonal on these wave packets, and its positivity gives the most stringent constraints. 

For simplicity, the ANEC operator can be placed at $x_2=(t,r,\vec{0}\,)$, which would correspond to putting a calorimeter in the `$x$' direction. We define light-cone coordinates for this point $x^{\pm}=t\pm r$.
Positivity of the ANEC operator on these states then reads
\begin{equation}
\langle\mathcal{E}\rangle=\lim_{r\rightarrow\infty}\, r^{d-2}\, \langle \mathcal{O}(q)^\dagger\,\int\limits_{-\infty}^\infty dx^-\,T_{--} (x^+,x^-)\,\mathcal{O}(q)\rangle \geq 0\, .
\end{equation}
Once expression \eqref{HM-state} for the states is introduced in the above and the position integrals have been pulled out of the correlator, we can use translation invariance to put the scalar operator to the right at $x_3=0$. The corresponding $d^dx_3$ integral becomes an overall factor, which will cancel with an identical factor when normalising $\langle\mathcal{E}\rangle$ with the 2-point function $\langle \mathcal{O}(q)\mathcal{O}(-q)\rangle$, see \eqref{2-pf-hofmal} (we will henceforth omit this factor). 
 With these manipulations, and assuming that the scalar operator is real, Hofman-Maldacena \cite{HofMal} claim that  
\begin{equation}
\langle \mathcal{E}\rangle=
\lim_{r\rightarrow\infty}\, r^{d-2}\, \int d^dx_1\, e^{i q t_1}\,\langle \mathcal{O}(x_1)\,\int\limits_{-\infty}^\infty dx^-\,T_{--} (x^+,x^-)\,\mathcal{O}(0)\rangle \, . \label{ANECexp}
\end{equation} 
evaluates the expectation value. As we shall see later, in momentum space it becomes clear that this quantity is indeed an expectation value and the $r \rightarrow \infty$ is in fact necessary. 
To compute this expectation value in momentum space, we note that the scalar operator to the left is effectively already in momentum space, with momentum $(q,\vec{0}\,)$. We further rewrite the other two operators in terms of their Fourier transforms. In particular, for the ANEC operator,
\begin{equation}
\int dx^- \int \frac{d^dp}{(2\pi)^d} \,e^{i(-p^{\scriptscriptstyle{0}} t+p^{1} r)}\,T_{--}(p)=2 \int \frac{d^d p}{(2\pi)^{d-1}}\,\delta(p^{\scriptscriptstyle{0}}+p^{\scriptscriptstyle{1}})\,e^{-i\frac{p^{\scriptscriptstyle{0}}-p^1}{2}x^+}\,T_{--}(p)\, ,
\end{equation}
where the Dirac $\delta$-function can be further integrated. Substituting the Fourier transforms in \eqref{ANECexp}, 
 \begin{align}
\langle \mathcal{E}\rangle=
 2\,\lim_{r\rightarrow\infty}\, r^{d-2}\,
\int\frac{d^{d-1}\vec{p}}{(2\pi)^{d-1}}\,e^{ 2i  p^1 r}
\llangle \mathcal{O}(q,\vec{0}\,)\,T_{--}(-p^1,\vec{p}\,)\,\mathcal{O}(p^1-q,-\vec{p}\,)\rrangle\, . \label{ANECexp2}
\end{align}
Using
\begin{equation}
 k_- = - \frac{1}{2} \,  k^+, \qquad \textrm{where} \quad k^+= k^{\scriptscriptstyle{0}} + k^{\scriptscriptstyle{1}}\, ,
\end{equation}
and expression \eqref{OTO-d} calculated in last section, the expectation value \eqref{ANECexp2} becomes
 \begin{align}\label{ANEC first integral}
\langle \mathcal{E}\rangle &=
 - \frac{4\, (d-1)\, c }{d-2}\lim_{r\rightarrow\infty}\, r^{d-2}\,
\int\frac{d^{d-1}\vec{p}}{(2\pi)^{d-1}}\,e^{2i p^1 r} \int \frac{d^dk}{(2\pi)^d}\,
\frac{\theta(q - k^{\scriptscriptstyle{0}}-|\vec{k}|)}{\,((q - k^{\scriptscriptstyle{0}})^2-|\vec{k}|^2)^{d -1- \Delta}} \notag \\[7pt]
& \hspace{95mm} \times \,  k^+  \,  \frac{\delta(k^{\scriptscriptstyle{0}} - |\vec{k}|)}{|\vec{k}|}  \,\delta\left(p^{\scriptscriptstyle{1}} + p^{\scriptscriptstyle{1}}_{\scriptscriptstyle{s}}\right),
\end{align}
where $c$ is defined in \eqref{cotodef}, and  $p^{\scriptscriptstyle{1}}_{\scriptscriptstyle{s}} = \frac{2 \, \hat{k}  \cdot \hat{p}  + |\hat{p} |^2 }{2 k^+} $. $\hat{k} $ is the $(d-2)$-dimensional transversal vector, $k=(k^{\scriptscriptstyle{0}}, \vec{k}\,)= (k^{\scriptscriptstyle{0}}, k^{\scriptscriptstyle{1}}, \hat{k} )$, and similarly for $p$. We have also used  
\begin{equation}
 \delta(k^2) = \frac{\delta(k^{\scriptscriptstyle{0}} - |\vec{k}|)}{2 \, |\vec{k}|} 
\end{equation}
(since  $k^{\scriptscriptstyle{0}} >0$), to simplify
\begin{equation}
 \theta(k^{\scriptscriptstyle{0}}- p^{\scriptscriptstyle{1}}) \, \delta((q - k^{\scriptscriptstyle{0}})^2-\big|\vec{k} + \vec{p} \, \big|^2) = \frac{1}{2 \, k^+}\delta\left(p^{\scriptscriptstyle{1}} + p^{\scriptscriptstyle{1}}_{\scriptscriptstyle{s}} \right)\, .
\end{equation}

The $p^{\scriptscriptstyle{1}}$ integral can now be evaluated with the Dirac $\delta$-function, leaving a $d^{d-2}\hat{p} $ integral. 
We next perform the change of variables
\begin{equation}
\hat{k} =\frac{\hat{l} }{x^+}-\frac{\hat{p} }{2}
\end{equation}
with which the factor $r^{d-2}$ in front of the integral is reabsorbed in the measure of the $d^{d-2}\hat{k} $, making the integral finite. With this change of variables $p^{\scriptscriptstyle{1}}_{\scriptscriptstyle{s}}\sim1/x^+$, hence goes to zero except in the exponential, which becomes
\begin{equation}
e^{-i\frac{\hat{p} \cdot \hat{l}}{k^{\scriptscriptstyle{+}}} }\, .
\end{equation}
The $d^{d-2}\hat{l} $ integral of the above exponential gives now a Dirac $\delta$-function $\delta^{(d-2)}(\hat{p} )$. Together with the previous one in \eqref{ANEC first integral}, which after the limit becomes $\delta(p^{\scriptscriptstyle{1}})$, these $d-1$ Dirac $\delta$-functions $\delta^{(d-1)}(\vec{p}\,)$ make explicit that $\langle \mathcal{E}\rangle$ is an expectation value, since they turn \eqref{ANECexp2}  into 
\begin{equation}
\langle \mathcal{E}\rangle\sim\llangle \mathcal{O}(q,\vec{0}\,)\,T_{--}(0,\vec{0}\,)\,\mathcal{O}(-q,\vec{0}\,)\rrangle\, . \label{exphofmal}
\end{equation}
The need for the $r\rightarrow\infty$ limit and the nature of this expectation value is therefore much more clear in momentum space. 

After the above Dirac $\delta$-functions have been performed, the only two remaining integrals are those of $dk^{\scriptscriptstyle{0}} dk^{\scriptscriptstyle{1}}$. The first can be performed with the remaining $\delta(k^{\scriptscriptstyle{0}} - |\vec{k}|)$, and the last integral can then simply be performed. 
All these steps make it clear that, unlike the calculation of $\langle \mathcal{E}\rangle$ in position space where one has to evaluate complicated integrals, in momentum space the expectation value reduces to integrals of $\delta$-functions and is hence much simpler.

The final result is
 \begin{align}
\langle \mathcal{E}\rangle &=
 - \frac{8\, (d-1)\, c }{(d-2) \, (2\, \pi)^{d+1}} \, q^{ \Delta+1-d} \int\limits_{0}^{q/2} dk^{\scriptscriptstyle{0}}\, ( k^{\scriptscriptstyle{0}})^{d-2} 
\,(q -2 \, k^{\scriptscriptstyle{0}})^{ \Delta+1-d} \notag \\[7pt] &= -  \frac{ 2^{ 1- 2 \Delta} \, \pi^{d/2+2}  \, \Gamma(d+1) \, a}{\Gamma(\Delta+1)\, \Gamma(\Delta+1 -d/2)\,\Gamma(d/2+1)^2}\,q^{2\,  \Delta+1-d} \, ,
\end{align}
where we have used equation \eqref{cotodef}.
We now normalise this expectation value with the 2-point function $\langle\mathcal{O}(q)\mathcal{O}(-q)\rangle$, with $\mathcal{O}$ given  again by \eqref{HM-state} (and as before, dropping the Gaussian factors), 
\begin{equation}
\langle\mathcal{O}(q)\mathcal{O}(-q)\rangle = \int d^dx_1 d^dx_3 \,e^{iq(t_1-t_3)} \langle \mathcal{O}(x_1)\mathcal{O}(x_3)\rangle\, .
\label{2-pf-hofmal}
\end{equation}
Using translation invariance to put $x_3=0$ in the correlator, we are left again with an overall $d^dx_3$ integral, which cancels the identical one in $ \langle \mathcal{E}\rangle$ in the numerator. Further writing the correlator in terms of its Fourier transform,
\begin{equation}
\langle\mathcal{O}(q)\mathcal{O}(-q)\rangle = \int d^dx_1  \,e^{iq t_1} \int\frac{d^dp}{(2\pi)^d} e^{ip\cdot x_1} \,N\,  G^{\Delta}(p)\, ,
\end{equation}
where $N$ is now the normalization of the 2-point function  (which is set to 1 in the rest of the paper). Introducing the expression for $G^{\Delta}(p)$ given by \eqref{2ptLm}, we obtain
\begin{equation}
\langle\mathcal{O}(q)\mathcal{O}(-q)\rangle =N\,\frac{\pi^{d/2+1}}{2^{2\Delta-d-1}\,\Gamma(\Delta)\,\Gamma(\Delta-d/2+1)} \,q^{2\Delta-d}
\end{equation}
where we have omitted the Heaviside step function since we have assumed $q>0$.

Using now
the relation between the normalization of the 2-point function $N$ and that of the 3-point function  $a$ which follows from Ward identities~\cite{OP},
\begin{equation}
a=-\frac{\Delta\,\Gamma(d/2+1)}{(d-1) \,\pi^{d/2}} \,N,
\end{equation}
we find
\begin{equation}
\frac{\langle \mathcal{E}\rangle}{\langle\mathcal{O}(q)\mathcal{O}(-q)\rangle}=\frac{q}{S_{d-2}}
\end{equation}
where 
\begin{equation}
 S_{d-2} = \frac{2\pi^{(d-1)/2}}{\Gamma(\frac{d-1}{2})}
\end{equation}
is the surface area of a $(d-2)$-sphere. This confirms the positivity of the ANEC expectation value.

\section{Discussion}
\label{sec:dis}

In this paper we study the 3-point function of Lorentzian CFTs in momentum space. We present the scalar 3-point function and give a prescription for calculating tensorial ones. The scalar 3-point function is presented both as an integral over an auxiliary momentum \eqref{eq1} and also as a triple-Bessel integral \eqref{3BL}. While the two expressions can be related via the analogous Euclidean expressions and we have checked that they are equivalent in a particular example, we have yet to understand the relation between them in Lorentzian space. Such a study could yield much simpler expressions for tensorial correlators and lead to a more  comprehensive understanding akin to the Euclidean case \cite{BMS:imp}. Furthermore, while we show how any tensorial 3-point correlator can be determined from the scalar 3-point function, the tensorial properties of the correlator may be obscured in the result. We hope to address these points in the future.

The $i\epsilon$ prescription in position space seems to indicate that renormalisation is not required in Lorentzian signature as opposed to Euclidean.
Our analysis in section \ref{sec:div} shows that our expression for the scalar 3-point function is consistent with this claim, since the cases which exhibit divergences in Euclidean signature do not arise in Lorentzian signature because of the different Bessel functions appearing. However, the issue of renormalisation in Lorentzian signature requires further investigation that we leave for future work.

We use our results to calculate the expectation value of the ANEC operator in states created by scalar operators. While this is the least interesting positivity result, simply stating that the state has positive energy, our calculation shows how natural such a computation is in momentum space compared to the position space calculation \cite{HofMal}. In particular, in momentum space the expectation value simply reduces to integrals over $\delta$-functions with the role of the $r \rightarrow \infty$ limit becoming much clearer---in this limit it is manifest that the transversal momenta of the ANEC operator vanish, \eqref{exphofmal}. This is convincing evidence that questions regarding ANEC may be more suitably framed in momentum space. By providing Lorentzian CFT correlators in momentum space, we hope that this is now possible. 

\section*{Acknowledgements}

We would like to thank Lorenzo Casarin, Chris Fewster, Thomas Hartman, Hugh Osborn, Guilherme Pimentel, Evgeny Skvortsov, Stefan Theisen and Alexander Zhiboedov for useful discussions.  We would like to thank the Mitchell Family Foundation for hospitality at the Brinsop Court workshop.  
\appendix

\section{Conventions and formulae}
\label{sec:not}

We use the $(-+++)$ signature. 
Our convention for the Fourier transform is
\begin{equation}
\tilde{f}(p)=\int d^d x \,e^{-i p\cdot x} \,f(x)\, ,\qquad f(x)=\int \frac{d^d p}{(2\pi)^d} \,e^{i p\cdot x} \,\tilde{f}(p)\, .
\end{equation}
The Dirac $\delta$-function is
\begin{equation}
\delta^{(d)}(p)=\int \frac{d^dx}{(2\pi)^d} \,e^{-i\,p\cdot x}\, .
\end{equation}
The Schwinger parametrisation formula is
\begin{equation}\label{schwinger}
\frac{1}{x^n}=\frac{1}{\Gamma(n)}\int\limits_0^\infty dt\,t^{n-1}\,e^{-t\, x}\, ,\qquad \text{when} \quad \Re(x)>0\, .
\end{equation}
The expansions of the Bessel functions are
\begin{align}\label{app:conv-Bessels}
J_\nu(z)&= \left(\frac{z}{2}\right)^\nu \sum\limits_{k=0}^\infty (-1)^k\,\frac{(z/2)^{2k}}{k!\,\Gamma(\nu+k+1)}\, ,\qquad\nu\neq -1, -2, -3, ... \nonumber\\
I_\nu(z)&=\left(\frac{z}{2}\right)^\nu \sum\limits_{k=0}^\infty \,\frac{(z/2)^{2k}}{k!\,\Gamma(\nu+k+1)}\, ,\qquad\nu\neq -1, -2, -3, ...\nonumber\\
K_\nu(z)&=\frac{\pi}{2\sin (\nu\pi)}\,\left( I_{-\nu}(z)-I_\nu(z)\right)\, , \qquad \nu\neq\mathbb{Z}\nonumber\\
Y_\nu(z)&=\frac{1}{\sin (\nu\pi)}\,\left(\cos(\nu\pi)\,J_\nu(z)-J_{-\nu}(z)\right), \qquad \nu\neq\mathbb{Z}.
\end{align}
In the large $z$ limit the Bessel functions can be approximated as
\begin{align}\label{Bessels-large-z}
J_\nu(z)\sim \sqrt{\frac{2}{\pi z}}\, \cos\left(z-\frac{\nu\pi}{2}-\frac{\pi}{4}\right) ,\quad 
K_\nu(z)\sim\sqrt{\frac{\pi}{2z}}\,e^{-z},\quad
Y_\nu(z)\sim\sqrt{\frac{2}{\pi z}}\, \sin\left(z-\frac{\nu\pi}{2}-\frac{\pi}{4}\right).
\end{align}

\section{3-point function of scalars from Fourier transform}
\label{sec:scalar-FT}

We compute the scalar 3-point function in momentum space by doing a direct Fourier transform of the position-space expression in four dimensions.\footnote{In general $d$ dimensions, the Fourier transform involves much more complicated integrals over the angles.}
Consider the Fourier transform of the Lorentzian correlator $\langle \mathcal{O}_1(x_1)\,\mathcal{O}_2(x_2)\,\mathcal{O}_3(x_3)\rangle$ for scalar operators with dimensions $\Delta_i$,  
\begin{equation} \label{3ptm}
\langle \mathcal{O}_1(p_1)\,\mathcal{O}_2(p_2)\,\mathcal{O}_3(p_3)\rangle=\int \prod_i d^4 x_i \,e^{- i \,p_i\cdot x_i}\,\frac{c_{123}}{(x_{23}^2)^{\beta_1}\,(x_{13}^2)^{\beta_2}\,(x_{12}^2)^{\beta_3}}\, ,
\end{equation}
where $\beta_j=\frac{\Delta_t}{2}- \Delta_j$, $\Delta_t=\D_1+\D_2+\D_3$,
 $x_{ij} = x_i - x_j$, and the $i\epsilon$ prescription dictates $t_j\rightarrow t_j-i\,\epsilon_j$ with $\epsilon_1>\epsilon_2>\epsilon_3$. Translation invariance allows us to strip off the momentum preserving $\delta$-function, which leads to 
\begin{equation}
\langle \mathcal{O}_1(p_1)\,\mathcal{O}_2(p_2)\,\mathcal{O}_3(p_3)\rangle= (2\pi)^4\, \delta^{(4)}(p_1+p_2+p_3)\, C(p_1,p_2; \{\beta_j\}) \,,
\end{equation}
with $C(p_1,p_2; \{\beta_j\})$ given by \eqref{3pt}, which we repeat here
\begin{equation}
 C(p_1, p_2; \{\beta_j\})  =  c_{123}\int \frac{d^4k}{(2\pi)^4} \int d^4 x_1 \, d^4x_2\, d^4x_3\,    \frac{ e^{- i (p_1-k) \cdot x_1} \,e^{- i (p_2+k) \cdot x_2} \,e^{- i k \cdot x_3}}{(x_{2}^2)^{\beta_1}\,(x_{1}^2)^{\beta_2}\,(x_{3}^2)^{\beta_3}}\, .
\end{equation}
This last expression is obtained by introducing a third coordinate $x_3$ via a Dirac $\delta$-function $\delta^{(4)}(x_3-x_{12})$, which is then reexpressed in terms of an additional $k$ integral. 

We now assume all $\beta_j>0$. In this case, we can use Schwinger parametrisation \eqref{schwinger} to exponentiate the distance factors in the denominator of the above integrand. For example, for the $x_1$ factor (we omit the subindex $1$ in the following expression to avoid cluttering),
\begin{equation}
\frac{1}{(x^2)^{\beta_2}}=\frac{1}{\Gamma(\beta_2)^2}\int\limits_0^\infty ds_1\,ds_2\, (s_1\,s_2)^{\beta_2-1} \,e^{-s_1(\epsilon_x +i \, (t_x +\lvert \vec{x}\rvert))}\,e^{-s_2(\epsilon_x +i  \,(t_x -|\vec{x} |))}\, ,
\end{equation}
and analogously for the $x_2$ and $x_3$ factors. We hence introduce a total of six Schwinger parameters $\{s_i\}$. When performing the Fourier transform in Euclidean signature, one can instead exponentiate the square of the distance factors $x_j^2$, thus requiring only three Schwinger parameters. This is possible because these factors are positive in Euclidean signature. In Lorentzian signature instead they need not be, and use of the Schwinger parametrisation is only allowed because the $i\epsilon_{ij}$  confer a positive real part to the distance factors. 

To perform the $x_1$ integral with the integrand above, we exchange it with the $s_1,s_2$ integrals. The two angles can then be integrated, and 
the integral over the time component $t_1$ gives $$\delta(k^{\scriptscriptstyle{0}}-p_1^{\scriptscriptstyle{0}}+s_1+s_2),$$ which can be used to integrate one of the Schwinger parameters.
To compute the radial integral, we first symmetrise its range of integration
by performing a change of variables on the remaining Schwinger parameter $s$ as
$u=s-\bar{u}$ with $\bar{u}\equiv (p_1^{\scriptscriptstyle{0}}-k^{\scriptscriptstyle{0}})/2$.
After doing the radial integral we obtain
\begin{equation}\label{u_1-integral}
\frac{(2\pi)^3\,\theta(p_1^{\scriptscriptstyle{0}}-k^{\scriptscriptstyle{0}})}{4\,\Gamma(\beta_2)^2\,|\vec{p}_1-\vec{k}|}
\int\limits_{-\bar{u}}^{\bar{u}} du\, \left({\bar{u}}^2-u^2\right)^{\beta_2-1}
\partial_{u}\Big( \delta(2u- |\vec{p}_1-\vec{k}|)-\delta(2u+ |\vec{p}_1-\vec{k}|) \Big)\, .
\end{equation}
Assuming $\beta_2\neq 1$, we do integration by parts and obtain
\begin{equation}
\frac{(2\pi)^3\,(\beta_2-1)}{4^{\beta_2-1}\,\Gamma(\beta_2)^2\,} 
\theta(p_1^{\scriptscriptstyle{0}}-k^{\scriptscriptstyle{0}}-|\vec{p}_1-\vec{k}|) \, 
|p_1-k|^{2\, \beta_2-4}\, ,
\end{equation}
where unambiguously $|p_1-k|=\sqrt{(p_1^{\scriptscriptstyle{0}}-k^{\scriptscriptstyle{0}})^2-|\vec{p}_1-\vec{k}|^2}$.
Following analogous steps for the $x_2$ and $x_3$ integrals, the  3-point correlator (when all $\beta_i\neq 1$) becomes
\begin{gather}\label{eq14d}
C(p_1,p_2; \{\beta_j\}))=  c(\{\beta_j\}) 
\int \frac{d^4k}{(2 \pi)^4}\,
\frac{\theta(p_2^{\scriptscriptstyle{0}}+k^{\scriptscriptstyle{0}}-|\vec{p}_2+\vec{k}|)\,
\theta(p_1^{\scriptscriptstyle{0}}-k^{\scriptscriptstyle{0}}-|\vec{p}_1-\vec{k}|)\,\theta(k^{\scriptscriptstyle{0}}-|\vec{k}|)
}{| p_2+k|^{4- 2\, \beta_1}\,|p_1-k|^{4- 2\, \beta_2}\,|k|^{4- 2\, \beta_3}} \, ,
\end{gather}
where 
\begin{equation}
c(\{\beta_j\}) =c_{123}\, (2\pi)^9\,\prod_i \frac{(\beta_i-1)}{4^{\beta_i-1}\,\Gamma(\beta_i)^2}\, .
\end{equation}
This precisely matches the 3-point function \eqref{eq1} calculated by Wick rotation in the case when $d=4.$ 

\section{3-point function for $\{d=3,\beta_j=1/2\}$ from the momentum-integrated expression}
\label{sec:app-check}

In this appendix we obtain the 3-point function of scalars with dimension $\Delta_j=1$, hence with $\beta_j=-\nu_j=1/2$, in $d=3$  from the momentum-integrated expression. This follows from substituting the corresponding 2-point function expression \eqref{k-integral-sp} in the 3-point function form
\eqref{3pt}, and reads (using $c_{123}=(2\pi)^{-3}$)
\begin{equation}
C(\{1/2\})= \int d^3 k\,
\frac{\delta(p_{2}^{\scriptscriptstyle{0}}+k^{\scriptscriptstyle{0}}-|\vec{p}_2+\vec{k}|)\,\delta(p_1^{\scriptscriptstyle{0}}-k^{\scriptscriptstyle{0}}-|\vec{p}_1-\vec{k}|)\,\delta(k^{\scriptscriptstyle{0}}-|\vec{k}|)}{|\vec{p}_2+\vec{k}|\,|\vec{p}_1-\vec{k}|\,|\vec{k}|}.
\end{equation}
The $k^{\scriptscriptstyle{0}}$ integral can readily be performed with the last Dirac $\delta$-function. Using
\begin{equation}
 \theta(p^{\scriptscriptstyle{0}})\delta(p^2)=\frac{\delta(p^{\scriptscriptstyle{0}}-|\vec{p}\,|)}{2 \,|\vec{p}\,|}
\end{equation}
for the remaining two Dirac $\delta$-functions, the integral becomes
\begin{equation}
C(\{1/2\})= 4 \int \frac{d^2 \vec{k}}{|\vec{k}|}\,
\theta(p_1^{\scriptscriptstyle{0}}-|\vec{k}|)\,\theta(p_2^{\scriptscriptstyle{0}} +|\vec{k}|)\,\delta\left((p_{2}+k)^2\right)\,\delta\left((p_1-k)^2\right)\biggr|_{k^{\scriptscriptstyle{0}}=|\vec{k}|}.
\end{equation}
The two step functions in the integrand confer an upper and lower bound on the radius $|\vec{k}|$, and further require $\theta(p_1^{\scriptscriptstyle{0}})\theta(p_1^{\scriptscriptstyle{0}}+p_2^{\scriptscriptstyle{0}})$, so the above becomes 
\begin{equation}\label{app-check-1}
4\,\theta(p_1^{\scriptscriptstyle{0}})\theta(-p_3^{\scriptscriptstyle{0}})\int\limits_0^{2\pi}d\varphi \left(
	 \theta(p_2^{\scriptscriptstyle{0}})	\int\limits^{p_1^{\scriptscriptstyle{0}}}_0 d|\vec{k}|\,
	+ \theta(-p_2^{\scriptscriptstyle{0}})   \int\limits^{p_1^{\scriptscriptstyle{0}}}_{-p_2^{\scriptscriptstyle{0}}} d|\vec{k}| \right)
	\,\delta\left((p_{2}+k)^2\right)\,\delta\left((p_1-k)^2\right)\biggr|_{k^{\scriptscriptstyle{0}}=|\vec{k}|}.
	\end{equation}
To compute the above integrals, it is much simpler to consider the case when $\vec{p}_1\cdot\vec{p}_2=|\vec{p}_1| \,|\vec{p}_2|$, i.e. when the two vectors are aligned and point in the same direction. The general case can then be obtained by an $SO(2)$ rotation of the spatial vectors in the final answer.

We first consider the integrand of the above two integrals. Performing the change of coordinates $\cos\varphi=\sigma$, it reads
\begin{align}
d\varphi\,\delta\left((p_{2}+k)^2\right)\,\delta&\left((p_1-k)^2\right)\biggr|_{k^{\scriptscriptstyle{0}}=|\vec{k}|} \nonumber\\[3mm]
&=
\frac{-d\sigma}{\sqrt{1-\sigma^2}}\,
\delta\left(p_2^2 +2|\vec{k}|(|\vec{p}_2|\sigma-p_2^{\scriptscriptstyle{0}})\right)
\delta\left(-p_1^2 +2|\vec{k}|(|\vec{p}_1|\sigma-p_1^{\scriptscriptstyle{0}})\right).
\end{align}	
(Henceforth we denote the radius as $k\equiv|\vec{k}|$ to avoid cluttering.)
We will use the second Dirac $\delta$-function to integrate the radius, so we rewrite it as
\begin{equation}	
	\delta\left(-p_1^2 +2k(|\vec{p}_1|\sigma-p_1^{\scriptscriptstyle{0}})\right)=\frac{\delta(k-\bar k(\sigma))}{2\bigl||\vec{p}_1|\sigma-p_1^{\scriptscriptstyle{0}}	\bigr|}, 
	\qquad \bar k(\sigma)=\frac{p_1^2}{2\left(|\vec{p}_1|\sigma-p_1^{\scriptscriptstyle{0}}\right)}.
\end{equation}	
We will then use the first Dirac $\delta$-function to integrate the angle, so we rewrite it as
\begin{equation}
	\delta\left(p_2^2 +2\bar k(\sigma)(|\vec{p}_2|\sigma-p_2^{\scriptscriptstyle{0}})\right)
	=\frac{\bigl||\vec{p}_1|\sigma-p_1^{\scriptscriptstyle{0}}\bigr|}{\bigl| p_2^2 |\vec{p}_1|+p_1^2|\vec{p}_2| \bigr|}\,\delta\left(\sigma-\bar\sigma\right),
	\qquad \bar\sigma=\frac{p_2^2 {p}_1^{\scriptscriptstyle{0}}+p_1^2 p_2^{\scriptscriptstyle{0}}}{p_2^2 |\vec{p}_1|+p_1^2|\vec{p}_2|}.
\end{equation}	
The remaining factor from the Jacobian becomes
\begin{equation}
\frac{1}{\sqrt{1-\bar\sigma^2}}
= \frac{\bigl| p_2^2 |\vec{p}_1|+p_1^2|\vec{p}_2| \bigr|}{|p_1|\,|p_2|\,|p_3|}
\end{equation}
where in the last step we have used the fact that $\sigma\in(-1,1)$, so the argument in the square root of the denominator is positive.\footnote{In doing the change of coordinates $\sigma=\cos\varphi$, we should in principle be careful with the integration range: the integrand $I(\varphi)$ does not satisfy $I(\varphi)=I(\varphi+\pi)$, therefore we should separate the integral over $\varphi\in(0,\pi)$ from the one over $\varphi(\pi,2\pi)$. Indeed, the two integrands become two different functions of $\sigma$, and the intermediate step functions are different, but the resulting integrated results are the same, so we need only multiply the result by a factor of 2.} 

Using the above equations, the first integral in \eqref{app-check-1} becomes
\begin{equation}
C_1=4 \,\frac{\theta(p_1^{\scriptscriptstyle{0}})\theta(p_2^{\scriptscriptstyle{0}})}{|p_1|\,|p_2|\,|p_3|}
\int\limits_{-1}^1 d\sigma \,\delta(\sigma-\bar\sigma)\,\int\limits^{p_1^{\scriptscriptstyle{0}}}_0 d k \,\delta(k-\bar k(\sigma))
\end{equation}	
The momentum integral gives a factor of $\theta(p_1^{\scriptscriptstyle{0}}-\bar k(\sigma))\theta(\bar k(\sigma))$. Using the expression for $\bar k$ above and doing some algebra, this becomes 
\begin{equation}
\theta(p_1^{\scriptscriptstyle{0}}-\bar k(\sigma)) \left[ \theta(p_1^{\scriptscriptstyle{0}}-|\vec{p}_1|)+\theta(|\vec{p}_1|\sigma-p_1^{\scriptscriptstyle{0}}) \right]=  \theta(p_1^{\scriptscriptstyle{0}}-|\vec{p}_1|).
\end{equation}
These step functions then do not constrain the range of integration of $\sigma$, whose integral then gives
\begin{equation}\label{app-check-2}
C_1=4 \,\frac{\theta(p_1^{\scriptscriptstyle{0}}-|\vec{p}_1|) \theta(p_2^{\scriptscriptstyle{0}})}{|p_1|\,|p_2|\,|p_3|}
	\theta(1-\bar\sigma)\theta(1+\bar\sigma).
\end{equation}
With some algebra it can be shown that
\begin{equation}\label{app-check-3}
\theta(1-\bar\sigma)\theta(1+\bar\sigma)=\theta(p_1^2\, p_2^2 \,p_3^2)=\theta(p_2^2)\theta(-p_3^2)+\theta(-p_2^2)\theta(p_3^2),
\end{equation}	
where in the last step we have used that the first step function in \eqref{app-check-2} implies $p_1^2<0$. Finally, taking into account the first two step functions in \eqref{app-check-2}, one can check that the second term in the above does not contribute, and the result becomes
\begin{equation}
C_1=4 \,\frac{\theta(p_1^{\scriptscriptstyle{0}}-|\vec{p}_1|) \theta(p_2^{\scriptscriptstyle{0}})\theta(-p_2^{\scriptscriptstyle{0}}+|\vec{p}_2|)\theta(-p_3^{\scriptscriptstyle{0}} -|\vec{p}_3|)}{|p_1|\,|p_2|\,|p_3|}.
\end{equation}
We now turn to the second integral in \eqref{app-check-1}, which reads
\begin{equation}
C_2=4 \,\frac{\theta(-p_2^{\scriptscriptstyle{0}})\theta(-p_3^{\scriptscriptstyle{0}})}{|p_1|\,|p_2|\,|p_3|}
\int\limits_{-1}^1 d\sigma \,\delta(\sigma-\bar\sigma)\,\int\limits^{p_1^{\scriptscriptstyle{0}}}_{-p_2^{\scriptscriptstyle{0}}} d k \,\delta(k-\bar k(\sigma)).
\end{equation}
The momentum integral gives now a factor of $\theta(p_1^{\scriptscriptstyle{0}}-\bar k(\sigma))\theta(p_2^{\scriptscriptstyle{0}}+\bar k(\sigma))$. As opposed to the first integral $C_1$, these step functions constrain now the range of $\sigma$, since they are equivalent to
\begin{equation}
\theta(p_1^{\scriptscriptstyle{0}}-|\vec{p}_1|)\,\theta(\sigma-\sigma_L)	,\qquad \sigma_L=\frac{2p_1^{\scriptscriptstyle{0}}p_2^{\scriptscriptstyle{0}}-p_1^2}{2p_2^{\scriptscriptstyle{0}}|\vec{p}_1|}.
\end{equation}
At this point the integral splits into two different integrals, depending on whether $\sigma_L$ is greater or smaller than the lower limit of integration $\sigma=-1$. The first integral brings about several additional step functions, namely $\theta(1-\sigma_L)\theta(\sigma_L+1)\theta(1-\bar \sigma)\theta(\bar\sigma-\sigma_L)$. It can be shown that these step functions together with the ones in front of the integral vanish. The remaining integral only requires the additional $\theta(-1-\sigma_L)=\theta(p_2^{\scriptscriptstyle{0}}-p_3^{\scriptscriptstyle{0}}-|\vec{p}_1|)$, and therefore becomes
\begin{equation}
C_2=4 \,\frac{\theta(p_1^{\scriptscriptstyle{0}}-|\vec{p}_1|)\theta(-p_2^{\scriptscriptstyle{0}})\theta(-p_3^{\scriptscriptstyle{0}})}{|p_1|\,|p_2|\,|p_3|}
\theta(p_2^{\scriptscriptstyle{0}}-p_3^{\scriptscriptstyle{0}}-|\vec{p}_1|)\,\theta(1-\bar\sigma)\theta(1+\bar\sigma).
\end{equation}	
We now use \eqref{app-check-3} above, from which only the first term contributes due to the four remaining step functions above, and the integral becomes
\begin{equation}
C_2=4 \,\frac{\theta(p_1^{\scriptscriptstyle{0}}-|\vec{p}_1|)\theta(-p_2^{\scriptscriptstyle{0}})\theta(p_2^{\scriptscriptstyle{0}}+|\vec{p}_2|)\theta(-p_3^{\scriptscriptstyle{0}}-|\vec{p}_3|)}{|p_1|\,|p_2|\,|p_3|}.
\end{equation} 	
Putting both $C_1$ and $C_2$ together, the 3-point function is finally
\begin{equation}
C(\{1/2\}) = \frac{4}{|p_1| \, |p_2| \, |p_3| } \theta(p^{\scriptscriptstyle{0}}_1 - |\vec{p}_1|)\,  \theta(- p^{\scriptscriptstyle{0}}_3 - |\vec{p}_3|)   \,\theta(p^{\scriptscriptstyle{0}}_2 + |\vec{p}_2|)  \, \theta(|\vec{p}_2|- p^{\scriptscriptstyle{0}}_2  ).
\end{equation}

\bibliographystyle{utphys}
\bibliography{ANEC}

\end{document}